\documentclass[preprint,NumberedRefs]{JASATesting}

\usepackage{etoolbox}
\usepackage{amsmath}
\usepackage{float}
\usepackage{booktabs}
\usepackage[utf8]{inputenc}
\usepackage{babel}
\usepackage{orcidlink}
\usepackage[margin=2.5cm]{geometry}
\usepackage{ifthen}
\usepackage{mathtools}
\newcommand{\summation}[3]{\displaystyle\sum_{#1=#2}^{#3}}

\newcommand{\infintegral}{\displaystyle\int_{-\infty}^{\infty}}
\newcommand{\symintegral}[1]{\displaystyle\int_{-#1}^{#1}}

\newcommand{\acos}{\operatorname{acos}}

\newcommand{\notetoself}[1]{\textcolor{red}{[#1]}}

\begin{document}
\title[CPSFM/spb]{The Chebyshev Polynomial Series Frequency Modulation Model for Waveform Design and Analysis}
\author{Stephen P.~Blackstock\orcidlink{0009-0008-4245-5528}}\email{sblackstock@utexas.edu}
\affiliation{Walker Department of Mechanical Engineering,  The University of Texas at Austin, Austin, TX 78712, USA} 

\author{Amaro Tuninetti\orcidlink{0000-0002-1805-1637}}\email{Amaro.Tuninetti@unh.edu}
\affiliation{Department of Biological Sciences,  University of New Hampshire, Durham, NH 03824, USA}
\author{Dieter Vanderelst\orcidlink{0000-0001-8049-5178}}\email{vanderdt@ucmail.uc.edu}
\affiliation{Department of Biological Sciences, University of Cincinnati,  Cincinnati, OH, USA} 
\author{Laura N.~Kloepper\orcidlink{0000-0003-4785-5279}}\email{laura.kloepper@unh.edu}
\affiliation{Department of Biological Sciences,  University of New Hampshire, Durham, NH 03824, USA}
\author{Michael R.~Haberman\orcidlink{0000-0002-7159-9773}}\email{haberman@utexas.edu}  
\affiliation{Walker Department of Mechanical Engineering,  The University of Texas at Austin, Austin, TX 78712, USA}
\date{\today} 

\begin{abstract}
Polynomial phase signals (PPS) are a staple of waveform design and analysis in sonar, radar, and communications fields. They also find application in the modeling of bioacoustic emissions,
especially those of echolocating animals such as bats and odontocetes. This work presents a novel PPS
waveform formulation that exploits some special properties of Chebyshev polynomials, such as orthogonality, recurrence relations, and equivalence to trigonometric functions. The result is the Chebyshev Polynomial Frequency Modulation (CPSFM) family of waveforms, which prove useful in the modeling of bioacoustic signals and the approximation of non-polynomial-phase signals such as hyperbolic chirps. We demonstrate that the CPSFM model admits compact analytic expressions for fundamental continuous-time signal processing functions such as the Fourier transform, the convolution and correlation operations, and the ambiguity function. Derivations for these expressions using CPSFM are presented, along with their application to the analysis of biosonar emissions of Mexican free-tailed bats.
\end{abstract}


\maketitle



\section{Introduction}
Frequency modulated (FM) waveforms are ubiquitous in engineering applications related to remote sensing technologies such as radar and sonar, and electromagnetic and acoustic communications\cite{Blunt_Mokole_2016}, in which waveform design is an engineering task whose goal is tactical performance and efficiency.  Common design considerations for practical applications are transmit power efficiency and total energy, autocorrelation characteristics, bandwidth, and Doppler sensitivity \cite{Cook_Bernfeld_1967}.  For communications applications, an additional consideration may be channel separation to enable efficient transmission and reception of multiple streams of information through the same medium with minimal interference, \textit{c.f.}~Davies et al.~and references therein\cite{davies2024}.
\\
FM signals are also evident in nature, where they are employed by animals for communication and echolocation \cite{Bradbury:2011aa}.  Obviously, animals do not engineer their waveforms the way humans do; rather, their tuned acoustic transmission signals are the result of millions of years of evolutionary experimentation \cite{simmons1980acoustic}. However, both engineering and evolution are driven by many of the same design considerations, resulting in strikingly similar outcomes for these two very different design pathways. It is for this reason that there has been significant interest in the study of biosonar emissions from a waveform engineering perspective.  An advanced understanding of the neurological, physiological, and behavioral aspects of animal echolocation can inform and support technological progress in remote sensing waveform design\cite{Vespe:2009aa}.\\ 
This work presents a waveform model, based on polynomial frequency modulation, which has relevance to both waveform design and the analysis of bioacoustic emissions. The model is simple, in the sense that it is characterized by a small number of parameters, but the waveform diversity that is representable by these parameters can be substantial.  Its development is motivated by our study of the echolocation signals of bats flying in dense swarms, where model simplicity enables the efficient characterization of millions of individual calls. 
\\
An FM waveform is characterized by an amplitude modulation function, $A(t)$, and a phase modulation function (PMF), $\varphi(t)$.   Mathematically, a waveform of this type is represented in the following analytic form\\
\begin{equation} \label{s_model}
	s(t) = A(t) \, e^{j\varphi(t)},
\end{equation}
where $\varphi(t)$ is proportional to the integral of the frequency modulation function (FMF), $f(t)$, of the waveform through the relation
\begin{equation} \label{phi_int_f}
	\varphi(t) = 2 \pi \int f(t) dt. 
\end{equation}\\
We are concerned here only with \textit{finite-duration} signals, where $A(t)$ has finite time-domain support.  Furthermore, we will restrict our discussion to \textit{constant-amplitude} and \textit{unit-energy} signals, such that  
\begin{equation} \label{unit_amplitude}
	A(t) = \frac{1}{\sqrt{T}} \operatorname{rect}{(\frac{t}{T})},
\end{equation}
where $T$ is signal duration.  We will refer to this unit-amplitude, finite-duration, frequency-modulated signal as an ``FM burst'', or simply an ``FM''.\\
 The parameters of $f(t)$ determine the distinctive qualities of FM signals, regardless of their origin (acoustic, electric, electromagnetic, or mechanical).  In acoustics, frequency modulation produces many familiar sounds, such as chirps, sirens, warbles, whistles, musical melodies, etc.\\ \\
The universe of FM waveforms may be divided into two broad mathematical categories:  those whose frequency modulation functions are algebraic, and those which are transcendental.  Algebraic FM waveforms include the widely used linear FM (LFM), the hyperbolic FM (HFM), and polynomial FMs.  Transcendental examples include the logarithmic FM, the exponential FM, and the sinusoidal FM (SFM).  The latter category also includes two interesting variations that were recently introduced by Hague and which have inspired the present work: the multi-tone sinusoidal FM (MTSFM)\cite{HagueDavidA.2021ATWD} and  the generalized sinusoidal FM (GSFM)\cite{HagueDavidA.2017TGSF}. The MTSFM family of waveforms employ a finite series of sinusoidal harmonics as the frequency modulation function, while the GSFM prescribes a frequency modulation function that is itself a linear FM waveform.  \\
Polynomial frequency modulation waveforms, commonly referred to as polynomial phase signals (PPS), find application in the analysis and modeling of FM signals whose frequency modulation function is not known, but for which spectro-temporal information is available.  Such a signal might be a simple LFM, such as a common active sonar transmission.  Or it could be a nonlinear FM, examples of which are the echo of an LFM off an accelerating target, a bird call, and even a whale song.  For nonlinear FM signals, we know from the Weierstrass approximation theorem that any frequency-modulated signal may be approximated, to any degree of precision desired, by a PPS \cite{PelegS.1991Eaco}. In this case, the parameters of the FMF are simply the coefficients of a polynomial of a finite degree.  And the parameters of the PMF, which are necessary for assembly of the waveform itself, are thus easily calculated through ploynomial integration, according to Eq.~\ref{phi_int_f}.  If a time-frequency distribution (TFD) of a signal, such as a spectrogram, is available, it may be analyzed to reveal the FMF parameters of the approximated waveform \cite{Djurovic:2017aa, GershmanA.B.2001Epom, PelegS.1991Eaco}, which may then be used to construct a noise-free prototype of a received signal, or to classify such a signal, or to explore signal transforms such as spectrum and correlation.
\\
Empirical estimation of the coefficients of a polynomial which defines the frequency modulation function of a PPS, given only a discrete TFD, is a somewhat hazardous task.  A naive approach involves sampling the peaks of the TFD, and then using linear-least-squares (LLS) or weighted-linear-least-squares (WLLS) method to find polynomial coefficients.  However, polynomial fitting via LLS or WLLS on uniformly-spaced data points is susceptible to numerical instabilities which may lead to poorer fits for higher-order polynomials.  A notable example of this problem involves Runge's Phenomenon \cite{Trefethen_2020}, where LLS-type fitting produces oscillations at the edges of the domain of the function to be approximated, and these oscillations grow exponentially with polynomial order.  There is a better way, which involves nonuniform sampling and a careful choice of polynomial basis functions.  The most notable example of this robust polynomial approximation of a function over a finite interval is Chebyshev approximation, which always leads to \textit{smaller} approximation error for higher-order polynomial fits \cite{Trefethen_2020}, provided that the function being approximated is analytic.
\\
The present work reformulates the modulation functions of PPS waveforms as Chebyshev polynomial series, and it introduces the Chebyshev polynomial series frequency modulation (CPSFM) class of waveforms.  It further provides the derivation of compact analytic expressions for spectrum, autocorrelation, cross-correlation, and ambiguity function of CPSFM waveforms, using expansions in generalized modified Bessel functions.  And it develops an approach to parameterization and analysis of such signals, which may be used for waveform design and analysis of bioacoustic signals, among other applications.\\
This work is motivated by a specific problem:  modeling the echolocation emissions of bats.  To that end, we seek a waveform model
\begin{enumerate}
\item which is convenient and efficient for fitting of an empirical time-frequency distribution of recorded emissions to a simple analytic expression, defined by a small number of parameters, for the frequency modulation function of a nonlinear FM,
\item whose parameters correspond to salient characteristics of the time-frequency representation of a waveform in an intuitive way, 
\item which admits closed-form solutions for analysis and signal processing tasks such as Fourier transform, autocorrelation, cross-correlation, convolution, Doppler processing, and ambiguity function,
\item which may be employed as an approximation, with any desired degree of accuracy, for an arbitrary FM waveform, thereby effectively transforming a waveform that does not conform to points (1), (2), or (3) into one that does.
\end{enumerate}
Point 3 above deserves some discussion.  Many signal processing functions take the form of integrals over time of the product of a time-domain signal and some other function of time.  It is very often the case that these integrals are difficult or impossible to evaluate analytically.  For example, if $s(t)$ is a unit-modulus PPS, its continuous Fourier transform, 
\begin{equation*}
	S(f) = \int_{-\infty}^{\infty} \exp{\left(j \sum_{n=0}^N c_n t^n \right)} e^{-j 2 \pi f t} dt	
\end{equation*}
involves Fresnel integrals for $N=2$ (a linear FM signal), and quickly becomes analytically intractable for higher order \cite{Scaglione1998}.  Were it available, an analytic, or mathematically exact, solution for the Fourier transform of a finite-duration PPS could be employed to, for example, efficiently derive the coefficients of a PPS with a prescribed spectrum.  Hence, a reformulation of the PPS model which admits analytic expressions for the Fourier transform and other signal processing functions is highly desirable.  Such a reformulation is described in section II.\\
The remainder of this paper is organized as follows.  Section II introduces the Chebyshev Polynomial Frequency Modulation waveform model, including a review of Chebyshev polynomials and series, along with the Jacobi-Anger expansion in Bessel functions, which plays an important role in derivations provided in Sec.~III.  Section III derives analytic solutions for common signal processing functions, previously elusive, involving polynomial-phase signals, where important equations are Fourier transform, Eq.~\ref{spectrum1}, ambiguity function, Eq.~\ref{AFboxed}, autocorrelation function, Eq.~\ref{ACFboxed}, and cross-correlation function, Eq.~\ref{CCFboxed}.  Section IV presents a CPSFM approximation of a hyperbolic FM signal and a CPSFM modeling of biosonar signals as two salient examples.  Finally, Sec.~ V provides a summary of the CPSFM waveform model and an outlook on its utility for signal analysis and waveform design.
\section{Chebyshev Polynomial Series Frequency Modulation (CPSFM)}
The proposed waveform model is based on Chebyshev polynomial series, which are well-known for their utility in the approximation of arbitrary functions \cite{Trefethen_2020}.
\subsection{Chebyshev polynomials}
Given that the waveform model developed in this work makes extensive use of Chebyshev polynomials, we begin with a short review of this unique polynomial and its relevant properties. 
The $n^\mathrm{th}$ Chebyshev polynomial is defined by 
\begin{equation}
	T_n(x) = \cos{ (n \operatorname{acos}{x}) }.
\end{equation}
In this work, it is convenient to make the substitution $x = \cos{\theta}$, which yields the following form for  the $n^\mathrm{th}$ order polynomial:
\begin{equation} \label{T_cos_theta}
	T_n(\cos{\theta}) = \cos{ n \theta }.
\end{equation}
The domain of a Chebyshev polynomial is generally restricted to $[-1, 1]$.  With this restriction, the set $\{T_0, T_1, T_2, ... T_N\}$ forms an orthonormal basis with weight $1 / \sqrt{1-x^2}$, so that 
\begin{align}
\begin{split}
	\symintegral{1} T_n(x) \, T_m(x)\frac{dx}{\sqrt{1-x^2}} &=
	\begin{cases}
		& 0 \quad m \ne n \\
		& \pi \quad m = n \ne 0 \\
		& \tfrac{\pi}{2} \quad m = n = 0 .
	\end{cases}
\end{split}
\end{align}
Within the domain of $[-1, 1]$, the range of a Chebyshev polynomial is also $[-1, 1]$. Within this domain and range, $T_n$ has exactly $n$ zeros - called \emph{Chebyshev nodes of the first kind}, and $n+1$ extrema ($\pm 1$), referred to as \emph{Chebyshev nodes of the second kind}.   These Chebyshev nodes, together with the orthogonality property, are central to the utility of Chebyshev polynomials in function approximation\cite{Trefethen_2020}.\\ [2.0ex]
Other properties of Chebyshev polynomials that are relevant in the present context include symmetry: $T_n(-x) = (-1)^n \, T_n(x)$, recurrence:  $T_{n+1}(x) = 2 x \,  T_n(x) - T_{n-1}(x) \text{ for } n > 0$, and integration: $\int T_n(x)\,dx = \left[(n-1) T_{n+1}(x) - (n+1) T_{n-1}\right]/\left[2(n^2-1)\right]$. \\
\subsection{Chebyshev polynomial series}
Since Chebyshev polynomials form an orthonormal basis, we may express an arbitrary function as a finite or infinite linear combination of $T_n(x)$, with the appropriate weights $a_n$. This linear combination is called a Chebyshev polynomial series (CPS), or simply Chebyshev series. In particular, any order-$N$ polynomial may be written as an order-$N$ Chebyshev series, provided the domain of the polynomial is restricted to a finite extent. This one-to-one mapping of polynomials to Chebyshev series is of critical importance to the mathematical derivations in this paper.\\
A Chebyshev polynomial series has the form
\begin{equation} \label{chebseries}
	\Phi^N(x; \{a_n\})=\summation{n}{0}{N} a_nT_n(x)
\end{equation}
where $N$ is the order of the series, \{$a_n$\} are the $N+1$ series coefficients $\{a_0, a_1, ..., a_N\}$, and $T_n(x)$ is the $n^\mathrm{th}$ Chebyshev polynomial.\\
Chebyshev series form a framework for function approximation that is robust and numerically stable, computationally simple and efficient, and in which coefficients are easlily intepreted \cite{Trefethen_2020}.  They also have the remarkable and extremely useful property that truncation of the higher-order terms results in a series which approximates the untruncated series better than any other polynomial of the truncated order.  That is, the best approximation of an order-$N$ Chebyshev series by an order-$(N-1)$ polynomial is the original Chebyshev series with the $N^\mathrm{th}$ coefficient set to zero \cite{boyd2001chebyshev}.  In contrast, if we express the same order-$N$ Chebyshev series as a simple order-$N$ polynomial, truncation to order-$(N-1)$ will generally result in a poor approximation of the original polynomial.\\
\\
With these definitions and properties in hand, we introduce a waveform model based on Chebyshev polynomial series, and we will rely on some specific properties of CPS.
\\
\\
First, the integral of an order-$N$ CPS is an order-$(N+1)$ CPS whose coefficients are determined from the coefficients of the integrand through the relation:
\begin{equation}
\begin{split}
	\int{\summation{n}{0}{N} a_nT_n(x) \, dx} = \alpha_0 + \summation{n}{1}{N+1} \alpha_n T_n(x)
\end{split}
\end{equation}
where $\alpha_0$ is an arbitrary integration constant, and the $\alpha_n$ coefficients are given by 
\begin{equation*}
	\quad \alpha_n = \frac{1}{2n}(a_{n-1} - a_{n+1}),
\end{equation*}
according to the integration property of Chebyshev polynomials.\\
Second, the sum of two CPS is clearly another CPS, whose vector of coefficients is simply the sum of the individual coefficients, i.e. given the following two CPS
\begin{equation*}
	\Phi_a(x) = \summation{n}{0}{N} a_nT_n(x) , \quad
	\Phi_b(x) = \summation{n}{0}{N} b_nT_n(x)
\end{equation*}
then
\begin{equation} \label{CPS_sum}
	\Phi_a(x) + \Phi_b(x)  = \summation{n}{0}{N} (a_n + b_n) T_n(x) \\
\end{equation}
Finally, and of primary importance to derivations herein: the domain of a CPS may be scaled and/or shifted to yield another CPS with different coefficients but same order, provided the modified domain is restricted to the interval $[-1, 1]$.  To wit, if
\begin{equation} \label{cps_scaleshift0}
	\Phi(\nu [x + \xi])=\summation{n}{0}{N} a_nT_n(\nu [x + \xi]),
\end{equation}
then
\begin{equation} \label{cps_scaleshift1}
	\Phi(\nu [x + \xi]) = \tilde{\Phi} (x)=\summation{n}{0}{N} \tilde{a}_{(\xi, \nu), n} \, T_n(x)
\end{equation}
where the modified coefficients $\{\tilde{a}_{(\xi, \nu), n}\}$ may be determined from the evaluation of $\Phi^N\left(\nu \left[x + \xi\right] \right)$ at the $N+1$ zeros $x_k$ of $T_{N+1}$, which are located at 
\begin{equation} 
	x_k = \cos{\left(\frac{k+\frac{1}{2}}{N+1} \, \pi \right)}, \quad k=\{0, 1,\ldots,N\} \, ,
\end{equation}
as follows:
\begin{equation} \label{shifted_coeffs}
\begin{split}
	\tilde{a}_{(\xi, \nu), 0} &= \frac{1}{N+1} \sum_{k=0}^{N} { \Phi (\nu [x_k+\xi])} \\
	\tilde{a}_{(\xi, \nu), n} &= \frac{2}{N+1} \sum_{k=0}^{N} { \Phi (\nu [x_k+\xi]) \, T_n(x_k)}.
\end{split}
\end{equation}
This scaling-and-shifting feature of Chebyshev series will be exploited to derive mathematically exact solutions to signal processing functions which involve time delay and time compression/expansion, such as the ambiguity function.
\subsection{Waveform model}
We now employ a CPS as the frequency modulation function of a generic PPS as an analytic signal with finite duration $T$ using the model provided in Eqs.~\ref{s_model},\ref{unit_amplitude} so that the waveform is nonzero only on the interval $\left[- \frac{T}{2} , \frac{T}{2} \right]$  and has unit energy.  Using Eq.~\ref{phi_int_f} , and given the restriction on $A(t)$, we see that $s(t)$ is uniquely determined, up to a constant offset in phase, by $f(t)$.  We call this model Chebyshev Polynomial Series Frequency Modulation, or CPSFM.
\\
\\
\subsubsection{CPSFM model}
As indicated above, we define the frequency modulation function of an order-$N$ CPSFM waveform to be an order-$(N-1)$ Chebyshev polynomial series:
\begin{equation} \label{CPSFM_fm}
	g(x)=\sum_{n=0}^{N-1} a_nT_n(x) \qquad -1 \le x \le 1 .
\end{equation}
Here, the first coefficient, $a_0$ determines the ``carrier'', or mean frequency, the second, $a_1$ is the bulk slope in time-frequency space, and $a_2 ... a_{N-1}$ enable the superposition of variations about the linear function to create an arbitrary polynomial FM. \\
\\
To accommodate a finite-duration waveform with an arbitrary time basis $\frac{-T}{2} \le t \le \frac{T}{2}$, we map that basis to $[-1,1]$ via:
\begin{equation} \label{t_mapping}
	x = \frac{2 t}{T}
\end{equation}
where $T$ and $t$ are the duration and time axis of the waveform in the natural time basis, respectively, and $x$ is the normalized time axis with duration 2 and limits $[-1,1]$. This normalization of time implies a normalization of frequency:  if the FMF of a waveform of duration $T$ has the form $f(t)$, the normalized FMF is $g(x=\frac{2t}{T}) = \frac{T}{2} f(t)$.  To emphasize that this mapping is a necessary condition for the employment of a Chebyshev polynomial basis, we will in the following mathematical expressions use the variables $x$ and $g$ to refer to normalized time and normalized frequency, respectively.\\
\\
Applying Eq.~\ref{phi_int_f} to Eq.~\ref{CPSFM_fm}, we obtain the waveform’s phase modulation function:
\begin{equation} \label{phasemodulation}
	\begin{split}
	\varphi(x) &= 2\pi \int{\sum_{n=0}^{N-1} a_nT_n(x)dt} \\
	&= \varphi_0 + \sum_{n=1}^N{\alpha_nT_n(x)},		
	\end{split}
\end{equation}
where $\varphi_0$ is an arbitrary integration constant which represents mean phase.  Finding the new coefficients $\alpha_n$ involves exploiting recurrence and integration relations of Chebyshev polynomials, which produces, for $n>0$
\begin{equation*}
	\quad \alpha_n = \frac{\pi}{n}(a_{n-1} - a_{n+1}).
\end{equation*}
\newline
If we assume, without loss of generality, zero mean phase ($\varphi_0=0$), we may write the CPSFM's time-domain waveform as
\begin{align} \label{tdw}
\begin{split} 
	s(x) &= \sqrt{2} \, \operatorname{rect} \left(\tfrac{x}{2}\right) \, e^{j\varphi(x)} \\ 
		&= \begin{cases}
			\sqrt{2} \, \exp{\left( j \sum_{n=1}^N \alpha_n T_n(x) \right)}, & -1 \le x \le 1 \\
			0, & \text{otherwise}.
		\end{cases} 
\end{split}
\end{align}
\\
Using the change of variables on which Eq.~\ref{T_cos_theta} is based, we have an alternative form of  Eq.~\ref{tdw}:
\begin{equation} \label{tdwcos}	
	s(\cos{\theta}; \{\alpha_n\}) = \exp{\left( j \sum_{n=1}^N \alpha_n \cos{n\theta}\right)} \qquad 0\le \theta \le \pi
\end{equation}
where $\{\alpha_n\}$ denotes the $N$-element vector $\{\alpha_1, \alpha_2, ...,\alpha_N\}$, which parameterizes the waveform function.  This formulation allows $s(x)$ to be expressed as an expansion in generalized Bessel funtions, which in turn facillitates the analytic evaluation of familiar signal processing functions which are defined as integrals of the product of $s(x)$ and some other function of $x$. These analytic solutions are the result of using a generalized Jacobi-Anger expansion of Eq.~\ref{tdwcos}, which is described in Sec.~\ref{sec:JA_exp}.\\
\subsection{Jacobi-Anger Expansion} \label{sec:JA_exp}
\subsubsection{Single variable}
The Jacobi-Anger expansion, which is a special case of Fourier sine or cosine series expansion \cite{kuklinski2021identities}, is given by the relation:
\begin{equation} \label{jax_1j}
	e^{j \alpha \cos{\theta}} = \sum_{m=-\infty}^{\infty}{j^m J_m(\alpha) e^{j m \theta}} = \sum_{m=-\infty}^{\infty} I_m(j \alpha) \, e^{jm\theta},
\end{equation}
where $J_m (\alpha)$ is the order-$m$ ordinary Bessel function of the first kind, and the far right-hand side expression is the order-$m$ modified Bessel function of the first kind, with argument $j \alpha$ via the identity: $I_m (j \alpha) \equiv (-j)^n J_m (-\alpha)$. The formulation provided in the second equality conveniently dismisses the $j^m$ oscillations, and provides a useful symmetry $I_{-m}(j \alpha) = I_m(j \alpha)$.
\\
The left-hand side of Eq.~\ref{jax_1j} is the trivial order-$1$ CPSFM whose phase modulation function is a first-order Chebyshev polynomial series. This waveform is a simple sinusoid in $x=\cos{\theta}$, with $\alpha$ dictating its frequency. Such a waveform is not generally considered frequency modulated, and so we are interested in CPSFM waveforms of order $N>1$, which requires the \textit{generalized} Jacobi-Anger expansion.
%
%
%
\subsubsection{Extension to multiple variable  }
The single-variable Jacobi-Anger expansion may be generalized by using a Fourier cosine series as the argument of the exponential:
\begin{equation} \label{jax_ni}
\begin{split}
	\exp{\left(j \sum_{n=1}^N \alpha_n \cos{n\theta}\right)} &= \sum_{m=-\infty}^{\infty} \mathcal{I}_m(j \{\alpha_n\}) \, e^{jm\theta}\\
\end{split}
\end{equation}
where $\{\alpha_n\}$ denotes the $N$-element vector $\{\alpha_1, \alpha_2, ..., \alpha_N\}$, and the coefficients $\mathcal{I}_m(j \{\alpha_n\})$ are the modified generalized Bessel functions (M-GBFs) evaluated at $j \{\alpha_n\}$\cite{Lorenzutta1995,kuklinski2021identities}.
This formulation satisfies the CPSFM model for any $N$, so Eq.~\ref{tdwcos} may be written as:
\begin{equation} \label{tdjax}	
	s(\cos{\theta}; \{\alpha_n\}) =  \sum_{m=-\infty}^{\infty} \mathcal{I}_m(j \{\alpha_n\}) \, e^{jm\theta} \qquad 0\le \theta \le \pi
\end{equation}
\\
The coefficients $\mathcal{I}_m(j \{\alpha_n\})$ may be calculated from the integral form of the M-GBF definition or from a recurrence formula which relates M-GBFs to GBFs (see Appendix \ref{appendix:M-GBF_calc}).
\\
This expansion of $s(\cos{\theta})$ in M-GBFs is the foundation for the derivation of analytic expressions for spectrum (Fourier transform), autocorrelation function, cross-correlation function, and ambiguity function of the CPSFM, as shown in Sec.~III.
\\
\section{Canonical signal analysis and processing functions}
The Chebyshev polynomial series frequency modulation waveform admits analytic solutions for Fourier transform, autocorrelation, cross-correlation, and ambiguity function, all of which are essential mathematical tools in the design of active radar and sonar waveforms, and which also prove useful in modeling of biosonar transmissions, as demonstrated in Sec.~\ref{eg_bats}. Detailed derivations of the following may be found in Appendix \ref{appendix:sp_derivations}.
\subsection{CPSFM Spectrum (Fourier transform)}
The continuous Fourier transform of a signal $s(t)$ is
\begin{equation} \label{FT0}
	S\big(f\big) = \mathcal{F}\{ s(t)\} = \infintegral s(t) \, e^{-j2 \pi f t} \, dt.\\
\end{equation}
For a finite-duration signal, such as $s(t) = \operatorname{rect}{(\frac{t}{T})}e^{j \varphi(t)}$, the spectrum may be written
\begin{equation} \label{FT1}
	S\big(f\big) = \int_{-T/2}^{T/2} s(t) \, e^{-j2 \pi f t} \, dt.\\
\end{equation}
If $s$ is a CPFSM waveform and we choose normalized time and frequency $x = \frac{2}{T} t$ and $g = \frac{T}{2} f$, so that $-1 \le x \le 1$, the resulting expression is
\begin{equation} \label{FT2}
\begin{split}
	S&\big(g; \{ \alpha_n\}\big) = \symintegral{1} \exp{\left( j \summation{n}{1}{N} \alpha_n T_n(x)\right)} \, e^{-j2 \pi g x} \, dx \\
	&= \int_{0}^{\pi} \exp{\left( j \summation{n}{1}{N} \alpha_n \cos{n \theta}\right)} \, e^{-j2 \pi g \cos{\theta}} \, \sin{\theta} d \theta, \\
\end{split}
\end{equation}
where the change of variables $x = \cos{\theta}$ is used.
Employing an expansion in generalized modified Bessel functions eventually yields
\begin{equation} \label{spectrum1}
	S\big(g; \{ \alpha_n\}\big) = \sum_{m=-\infty}^{\infty} \gamma_{m} \; \mathcal{I}_m(j \{\hat{\alpha}_n (g) \}),
\end{equation}
where
\begin{align*}
\begin{split} 
	\gamma_{m}
		&= \begin{cases}
			\frac{2}{1 - m^2}, &  m  \operatorname{even} \\
			\pm \frac{j \pi}{2}, &   m  = \pm 1 \\
			0, &  m \text{ otherwise odd}
		\end{cases} \\
	\hat{\alpha}_n(g)
		&= \begin{cases}
			\alpha_n - j 2 \pi g, & n=1 \\
			\alpha_n, & n>1 .
		\end{cases} 
\end{split}
\end{align*}
Thus, the spectrum, or continuous Fourier transform, of an order $N$ CPSFM is uniquely determined, in analytic form, by its $N$ coefficients $\{a_0, a_1, ..., a_{N-1}\}$ and duration $T$, which specifies the mapping to normalized time and frequency.  This result extends to any finite-duration polynomial-phase signal, since every polynomial may be expressed as a Chebyshev series.  To the authors' knowledge, an analytic solution to the Fourier transform of a PPS of arbitrary order has not been previously been offered.
\subsection{Ambiguity function of the CPSFM}
The \textit{ambiguity function} (AF) of a waveform is a powerful tool for evaluating both range resolution performance and Doppler sensitivity of transmit waveforms in sonar and radar applications \cite{Ricker:2012}.  For narrowband waveforms, or when the relative velocity between observer and target is very small relative to wave propagation speed, the AF is usually approximated by the narrow-band AF, which admits a \textit{frequency shift} but not \textit{frequency scaling}.  In sonar applications, object velocities are generally a small but significant fraction of propagation speed, and bandwidth can be an octave or more.  In this case, the Doppler effect on frequency cannot be well-approximated by a simple scalar shift.  Rather, it is exactly indicated by stretching or compressing the frequency axis, which is incorporated into the wideband AF.  We will limit our treatment here to the wideband version. 
\\
The wideband ambiguity function (AF) of a finite-duration signal $s(t)$ is00
\begin{equation} \label{AF_definition}
\begin{split}
	\chi(\tau, \nu) &= \frac{\sqrt{\nu}}{T} \int_{-\frac{T}{2}}^{\frac{T}{2}} s^*(t) \, s(\nu (t + \tau)) dt ,\\
\end{split}
\end{equation}
where $\nu = \frac{1+\mu}{1-\mu}$, $\mu = \frac{v}{c}$ is Mach index, $c$ is sound speed, and $v$ is relative velocity.  Normalizing by $\frac{\sqrt{\nu}}{T}$, where $T$ is the waveform duration, enforces $|\chi(\tau, \nu)| \le 1$. \\
\\
If $s$ is a CPSFM, then we use normalized time  $x = \frac{2 t}{T}$ and delay $\xi = \frac{2 \tau}{T}$, so
\begin{equation} \label{AF_CPSFM_0}
\begin{split}	
	\chi(\xi, \nu) &= \frac{\sqrt{\nu}}{2} \int_{x_1}^{x_2} s^*(x) \, s(\nu (x + \xi)) dx ,\\
\end{split}
\end{equation}
where the limits $x_1$ and $x_2$ satisfy the restrictions $-1 \le x \le 1$ and $-1 \le \nu(x + \xi) \le 1$.\\
Expressing $s(x)$ as a CPSFM, we have
\begin{equation}
\begin{split}
	\chi&(\xi, \nu ; \{ \alpha_n\}) = \rho_{\xi, \nu} \, \frac{\sqrt{\nu}}{2} \\
	& \times \int_{x_1}^{x_2}  \, \exp{ \left( j \sum_{n=1}^{N} \big( \tilde{\alpha}_{(\xi, \nu),n} - \alpha_n \big) \, T_n(x)\right)} dt ,
\end{split}
\end{equation}
where $\{\tilde{\alpha}_n\}$ are the Chebyshev series coefficients modified by a scaling and shifting of the argument, as indicated by Eq.~\ref{cps_scaleshift1}, and
 $\rho_{\xi, \nu} = \exp{ \left( j \tilde{\alpha}_{(\xi, \nu),0} \right)}$, which represents an offset in mean phase imposed by $\xi$-shifting and $\nu$-scaling, and is generally nonzero for $\xi \ne 0$ and $\nu \ne 1$. \\
\\With a change of variables: $x = \cos{\theta}, \quad T_n(\cos{\theta}) = \cos{n\theta}$, and an expansion in M-GBFs, we arrive at
\begin{equation} \label{AFboxed}
\begin{split}
	\chi(\xi, \nu ; \{ \alpha_n\}) &= \rho_{\xi, \nu} \frac{\sqrt{\nu}}{2} \\
	&\times \sum_{m=-\infty}^{\infty} \gamma_m(\theta_1, \theta_2) \, \mathcal{I}_m \Big(j \{ \tilde{\alpha}_{(\xi, \nu), n} - \alpha_n\} \Big) ,
\end{split}
\end{equation}
where
\begin{equation} \label{gammaplusdef}
\begin{split}
	\gamma_m&(\theta_1, \theta_2) = \frac{1}{m^2-1}
	\Big[ \Upsilon_m(\theta_1) - \Upsilon_m(\theta_2)\Big] \quad , m \ne \pm 1\\
	\gamma&_{\pm 1}(\theta_1, \theta_2) = \frac{1}{4} \Big[ e^{j 2 m \theta_2} - e^{j 2 m \theta_1} + j 2 m (\theta_1 - \theta_2) \Big] \\
	\theta_{i} &= \acos{x_{i}} \, , \, i \in \{1,2\}\\
	x_1 &= \max{(-1, -\frac{1}{\nu} -\xi)} \\
	x_2 &= \min{(1, \frac{1}{\nu} -\xi)} \\
	\Upsilon_m (\theta) &= e^{j m \theta} ( \cos{\theta} - j m \sin{\theta}).
\end{split}
\end{equation}
\\
The wideband ambiguity function is actually a special case of the more general cross-ambiguity function (CAF), which provides a visualization of how the cross-correlation of two signals is affected by Doppler.  The CAF has the form
\begin{equation} \label{CAF_definition}
\begin{split}
	\chi_{ab}(\tau, \nu) &= \frac{\sqrt{\nu}}{T} \int_{-\infty}^{\infty} s_a^*(t) \, s_b(\nu [t + \tau]) dt ,\\
\end{split}
\end{equation}
where $s_a$ and $s_b$ are two distinct time-domain signals.  If these signals are CPSFM waveforms of identical duration whose phase modulation functions take the form
\begin{equation*}
\begin{split} 
	\varphi_a(x) &=   \sum_{n=1}^N \alpha_n T_n(x) \\
	\varphi_b(x) &=   \sum_{n=1}^N \beta_n T_n(x), 
\end{split}
\end{equation*}
the CAF derivation procedes just as that for the AF does, and yields
\begin{equation} \label{CAFboxed}
\begin{split}
	\chi_{ab}(\xi, \nu ; \{ \alpha_n\}) &= \rho_{\xi, \nu} \frac{\sqrt{\nu}}{2} \\
	&\times \sum_{m=-\infty}^{\infty} \gamma_m(\theta_1, \theta_2) \, \mathcal{I}_m \Big(j \{ \tilde{\beta}_{(\xi, \nu), n} - \alpha_n\} \Big), 
\end{split}
\end{equation}
where the parameters $\gamma_m, \theta_{i},  x_{i}$, $i \in \{1,2\}$, are as specified in Eq.~\ref{gammaplusdef}, but $\rho_{\xi, \nu}$ is now $\rho_{\xi, \nu} = \exp{ \left( j \tilde{\beta}_{(\xi, \nu),0} \right)}$.
\\
These analytic expressions for AF and CAF set the stage for two special cases:  the autocorrelation function (ACF) and the cross-correlation function (CCF).
\\
\subsection{CPSFM Autocorrelation and Cross-correlation}
Autocorrelation is a widely-used signal processing tool which, among other things, gives a succint characterization of the performance of an active sonar or radar which employs a given signal as its transmit waveform.  It ``compresses'' the energy of a waveform into its most compact form, which provides information about range resolution and uncertainty, under ideal conditions, in pulse-compression-based remote sensing systems \cite{Ricker:2012}.  We are therefore interested in an analytic solution for the autocorrelation function of a CPSFM, which makes efficient the evaluation of how different sets of CPSFM parameters lead to differences in measures of performance, such as main-lobe-to-side-lobe ratio.\\
The autocorrelation function of a constant-modulus, finite-duration signal $s(t)$ is:
\begin{equation} \label{ACF_definition}
\begin{split}
	R(\tau) &= \frac{1}{T} \int_{-\frac{T}{2}} ^{\frac{T}{2}} \, s^*(t) s(t + \tau) \, dt ,
\end{split}
\end{equation}
where $T$ is the waveform duration whose reciprocal imposes the condition $|R(\tau)| \le 1$. \\
This expression is the same as Eq.~\ref{AF_definition}, with $\nu=1$, so the CPSFM ACF is
\begin{equation} \label{ACFboxed}
	R(\xi ; \{ \alpha_n\}) = \frac{\rho_\xi}{2} \sum_{m=-\infty}^{\infty} \gamma_m(\theta_1, \theta_2) \, \mathcal{I}_m\Big( j\{\tilde{\alpha}_{\xi,n} - \alpha_n \}\Big).
\end{equation}
\\
%
Cross-correlation is the more versatile cousin of autocorrelation.  It provides an efficient mechanism for evaluating waveform performance under conditions which are \emph{not} ideal.  These conditions include noise, channel characteristics such as multipath and spectral shaping, and the presence of competing transmissions which overlap spectrally.  The cross-correlation function is an essential tool for the design of active sonar and radar waveforms and the analysis of passively-recorded signals such as biosonar emissions.\\
The CCF of two signals $s_a(t)$ and $s_b(t)$, of identical duration $T$, is
\begin{equation*}
	R_{ab}(\tau) = \frac{1}{T} \int_{-\infty}^{\infty} { s^*_a(t) \, s_b(t + \tau) \, dt} \, . \\
\end{equation*}
This expression is the same as Eq.~\ref{CAF_definition}, with $\nu=1$, so the CPSFM CCF is
\begin{equation} \label{CCFboxed}
	R_{ab}(\xi ; \{ \alpha_n, \beta_n \}) = \frac{\rho_\xi}{2} \sum_{m=-\infty}^{\infty} \gamma_m(\theta_1, \theta_2) \, \mathcal{I}_m\Big( j\{\tilde{\beta}_{\xi,n} - \alpha_n \}\Big).
\end{equation}
\section{Selected CPSFM applications}
\ifdefined\HideExamples
\notetoself{Examples hidden}
\else
\subsection{Example 1: CPSFM approximation of hyperbolic FM}
The hyperbolic FM (HFM) waveform is well-known in sonar and radar as the optimally ``Doppler-tolerant'' waveform, meaning that its matched-filtered response incurs minimum degradation in range resolution in the face of nonzero Doppler\citep{YangJ.2006Dpoh}. The frequency modulation function of an HFM is
\begin{equation} \label{HFM_fm}
	 f(t) = \frac{f_1 \, f_2 \, T}{(f_1 - f_2) \, t +f_2 \, T}, 
\end{equation}
where $f_1$ and $f_2$ are initial and terminal frequencies, and $T$ is waveform duration.
This function, while not transcendental, cannot be expressed as a finite-length polynomial.  It can, however, be approximated, to any desired accuracy, by a Chebyshev polynomial series.  Such an approximation therefore admits a closed-form solution for the AF, which provides a metric for evaluating the trade-off between the approximation's CPSFM order and its Doppler tolerance.  \\
Figure \ref{CPSFM_Fig_1} shows AFs of an HFM, and also low-order approximations by CPSFM.  Note that the $N=2$ case is simply a linear FM (LFM) whose slope, or chirp rate, is the average slope of the HFM.  The LFM clearly has a substantially different Doppler tolerance profile, while the $N=4$ rendition is very similar to the HFM.
\begin{figure*}[h!] 
\figline{ \fig{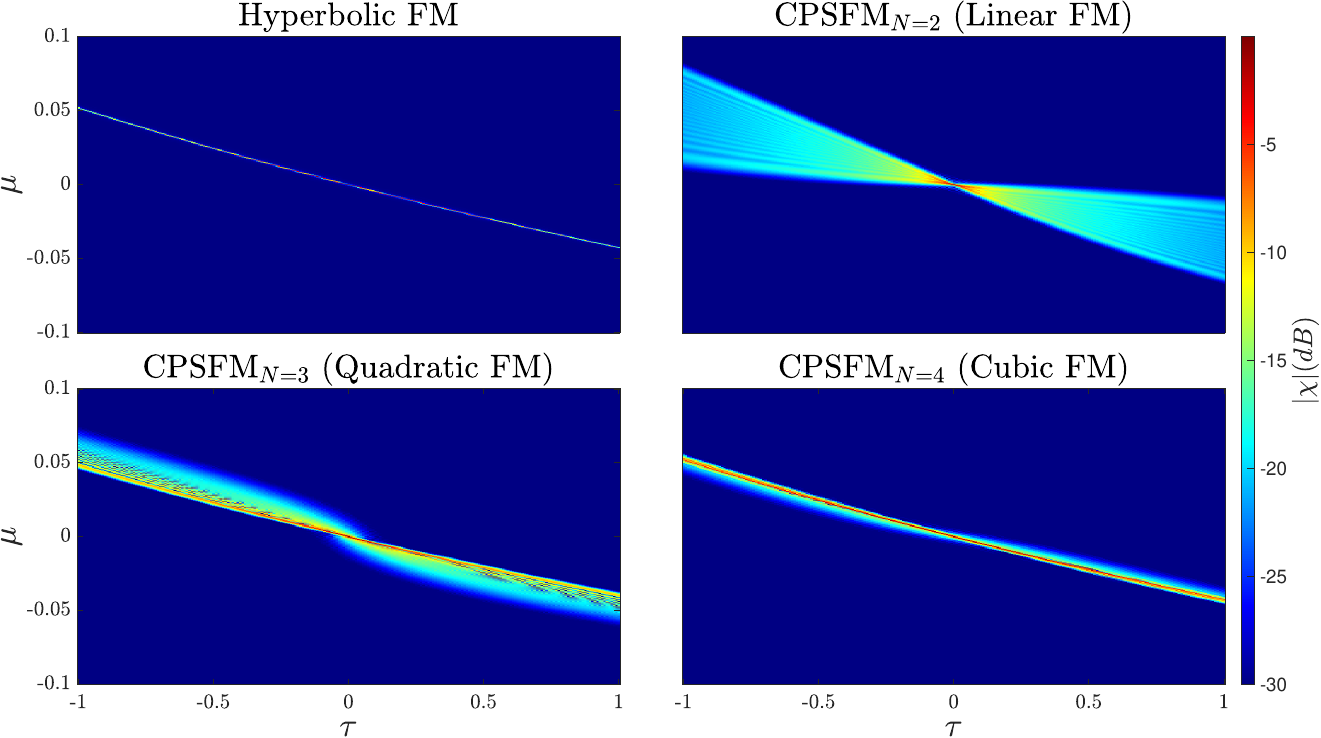} {16cm} {} \label{CPSFM_Fig_1}} 
\caption{Ambiguity function for HFM waveform and 3 CPSFM approximations.  The second-order CPSFM, which is an LFM, performs poorly, with substantially degraded SNR and range resolution characteristics.  The fourth-order CPSFM shows good Doppler tolerance.}
\end{figure*} \\
\begin{figure*} [h!] 
\figline{ \fig{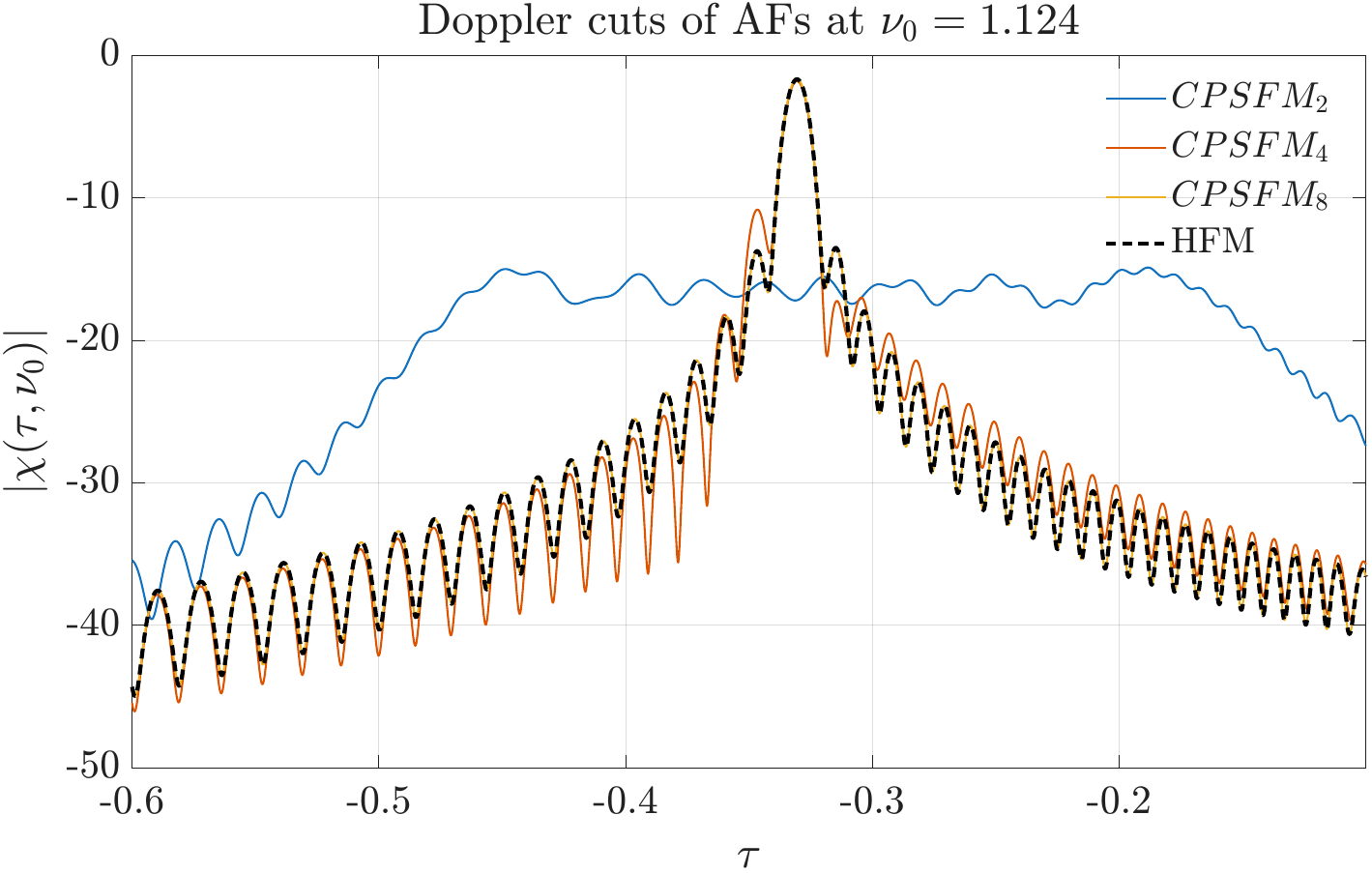} {9cm} {} \label{CPSFM_Fig_2}}
\caption{AF Doppler cuts at $\nu_0=1.124$ for true HFM and three CPSFM approximations.  Initial and terminal frequencies are 100kHz and 200kHz, and duration is 2ms.  The chosen Doppler coefficient corresponds to relative velocity of 20m/s.}
\end{figure*}
To quantify the Doppler tolerance characteristics of the CPSFM approximations, we may produce ``Doppler cuts'' of the AF, which present a cross-correlation between a waveform and its Doppler-warped self, for a specific relative target velocity.  
Figure \ref{CPSFM_Fig_2} shows the Doppler cuts, based on a relative velocity of 20m/s in air, of a few CPSFM approximations of the HFM waveform.  The $N=2$ CPSFM shows a substantially degraded pulse compression output (broader and smaller amplitude) compared to the HFM, while the $N=8$ CPSFM is virtually identical to the HFM.  For almost all practical applications, $N=8$ is clearly a very good substitute for the actual HFM, and the $N=4$ CPSFM would suffice for some applications.\\
It is unlikely that an approximation of an HFM waveform would be viewed as superior in some way to a digitally-sampled HFM in modern sonar and radar transmit systems.  However, this rather academic example provides some insight into how the CPSFM model might be applied to transmit waveform design.
\subsection{Example 2: biosonar vocalizations of Mexican free-tailed bats} \label{eg_bats}
The subject of the preceding example was waveform \textit{design}.  In this example, the CPSFM model will be applied to the problem of waveform \textit{analysis}.\\
An interesting use of CPSFM waveforms, and the guiding motivation of this paper, is the modeling of biosonar emissions.  Mexican free-tailed bats (\textit{Tadarida brasiliensis}) live in large colonies, sometimes measured in the millions of individuals.  Every evening, they emerge from their roosts - a cave, or an urban bridge - to feed on insects that they acquire through the use of sonar.  During this emergence, the swarms are dense, and the bats appear to use sonar to avoid colliding with each other and with other obstacles (such as, for example, a hawk \cite{kloepper2024hawkear}).
\\
Figure \ref{CPSFM_Fig_3} shows spectrograms of two individual bat calls (amidst many other bat calls) from a recording of a dense swarm, emerging from the Jordana Caves, in Sierra County, NM, in July 2022, at dusk.  These spectrograms exhibit time-frequency ``ridges'', which are the spectrogram's manifestation of instantaneous frequency of signal energy.  In our treatment, these ridges are sampled in time and frequency, using a lightly-supervised Canny edge detector \cite{Canny1986}, and the samples are then fit to a fourth-order Chebyshev polynomial series via weighted linear least squares.  The choice of fourth order is due to visual inspection of the ridge structure for over 1000 isolated calls, all of which exhibit at most one critical point and at most two inflection points.  The minimum polynomial order for such a function is four.\\
\begin{figure*}[h!]
\figline{ \fig{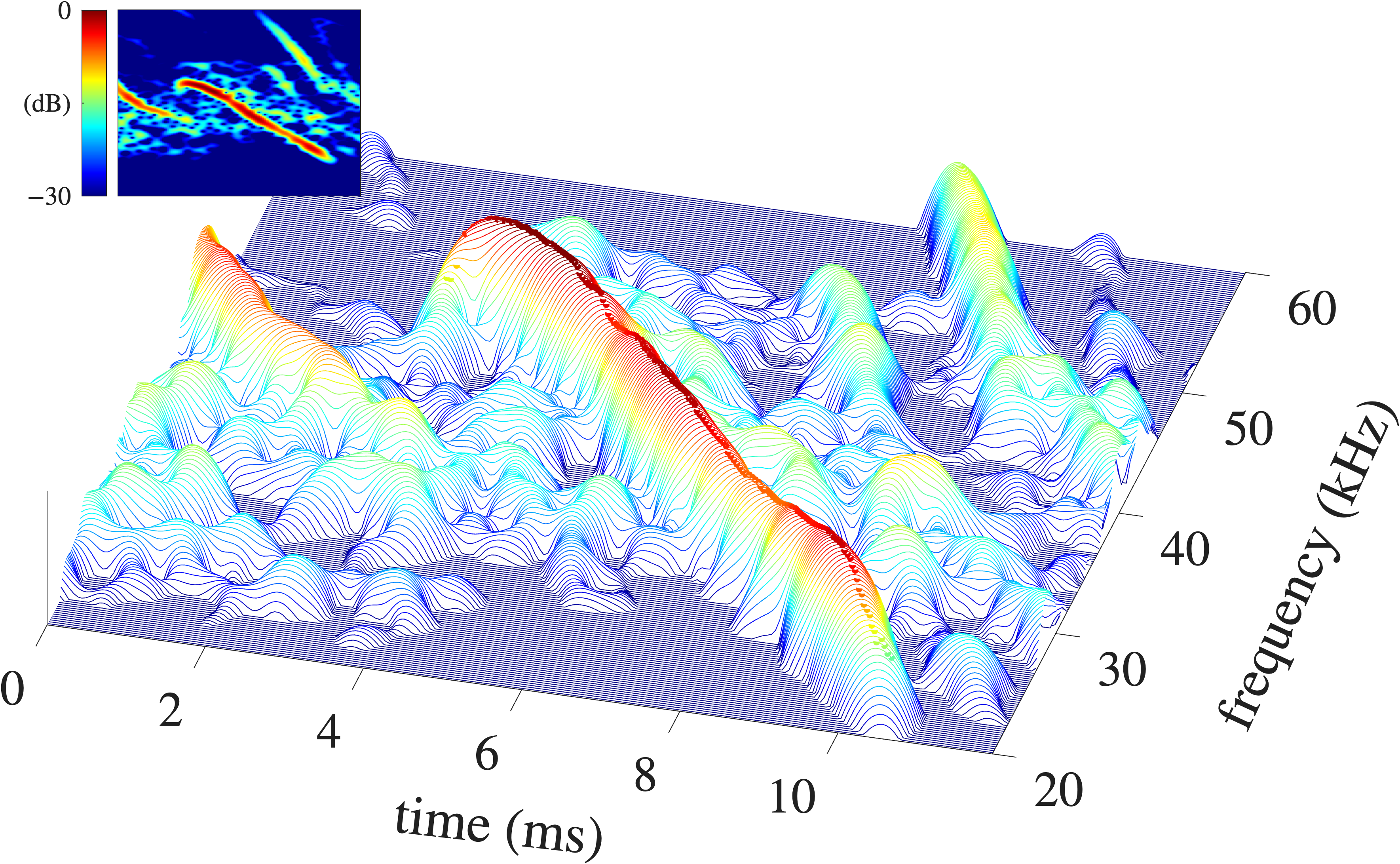} {8cm} {(a)}
	\fig{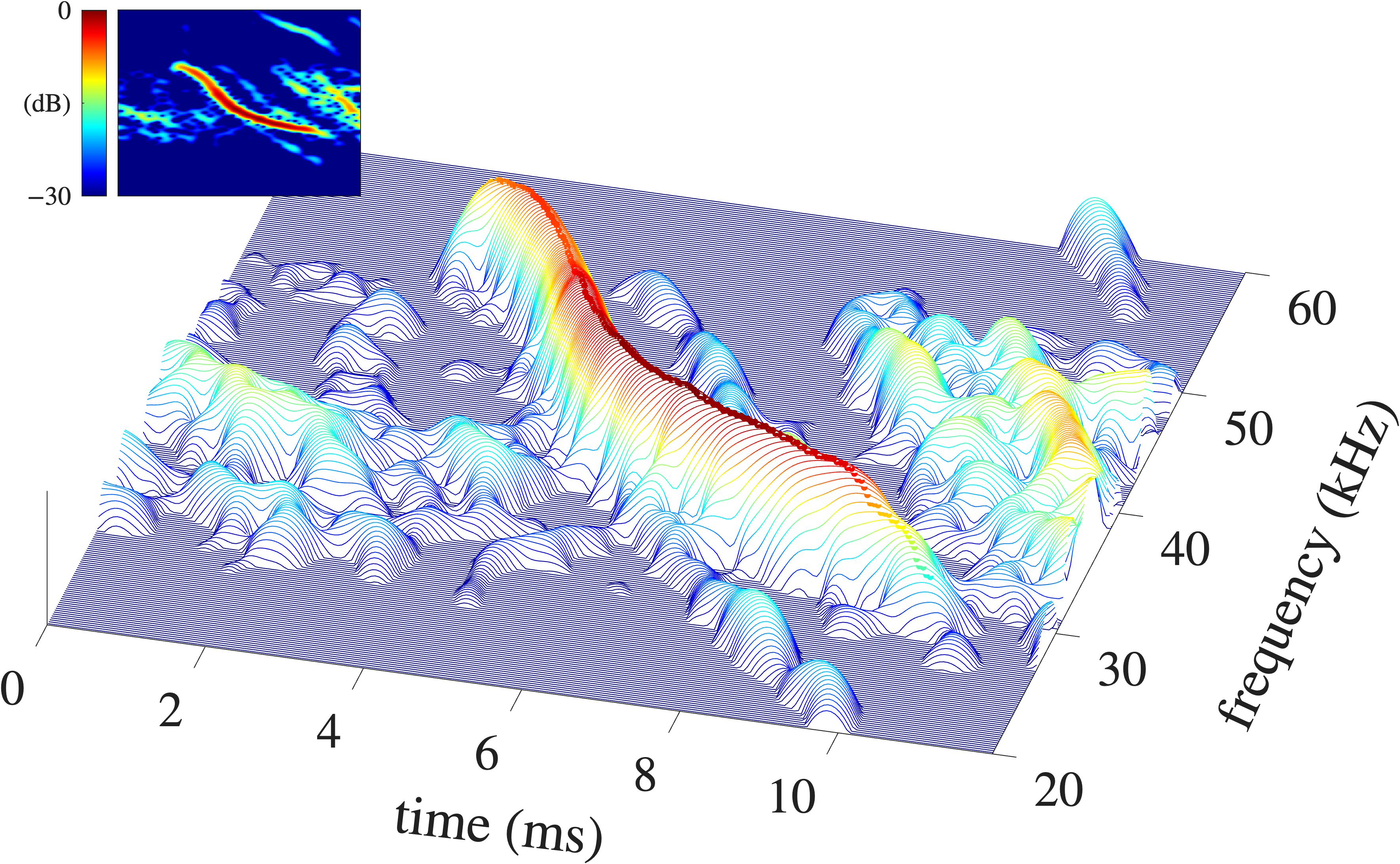} {8cm} {(b)}}
\caption{Spectrograms of two segments (a) and (b) of a recording of the echolocation transmissions of a swarm of \textit{Tadarida brasiliensis}.  Note that the axes are un-normalized time and frequency $\{t, f\}$, as opposed to normalized $\{x, g\}$, to make the figures easier to interpret. \label{CPSFM_Fig_3}
}
\end{figure*}\\
Figure \ref{CPSFM_Fig_4} shows the modeled instantaneous frequency functions for the CPSFM fits, and also their spectra, calculated according to Eq.~\ref{spectrum1}, where the M-GBF expansion was truncated at $M=1000$. This limit was more than sufficient to reduce errors to machine precision.
\\
\begin{figure*}[h!]
\figline{ \fig{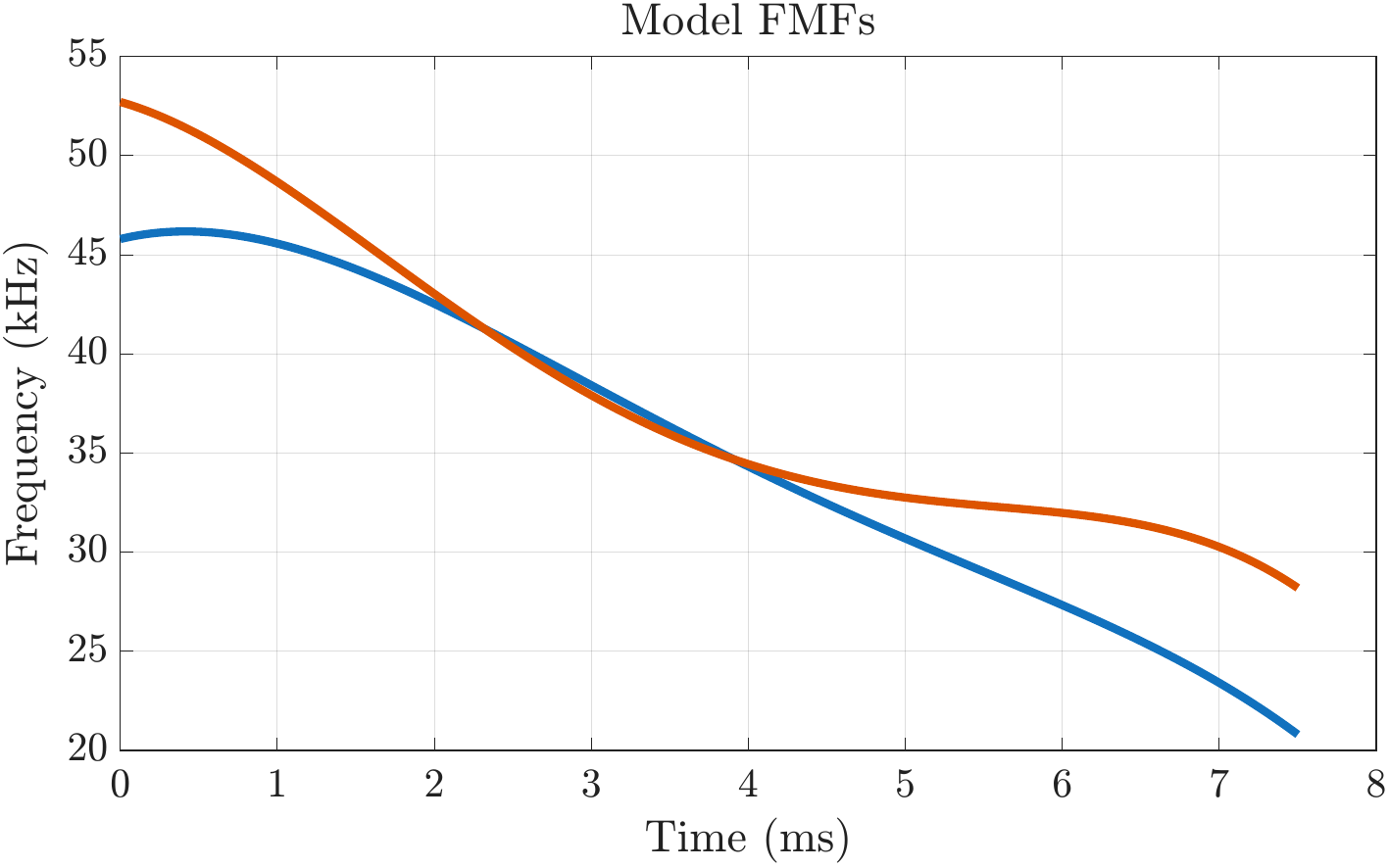} {8cm} {(a)}
	\fig{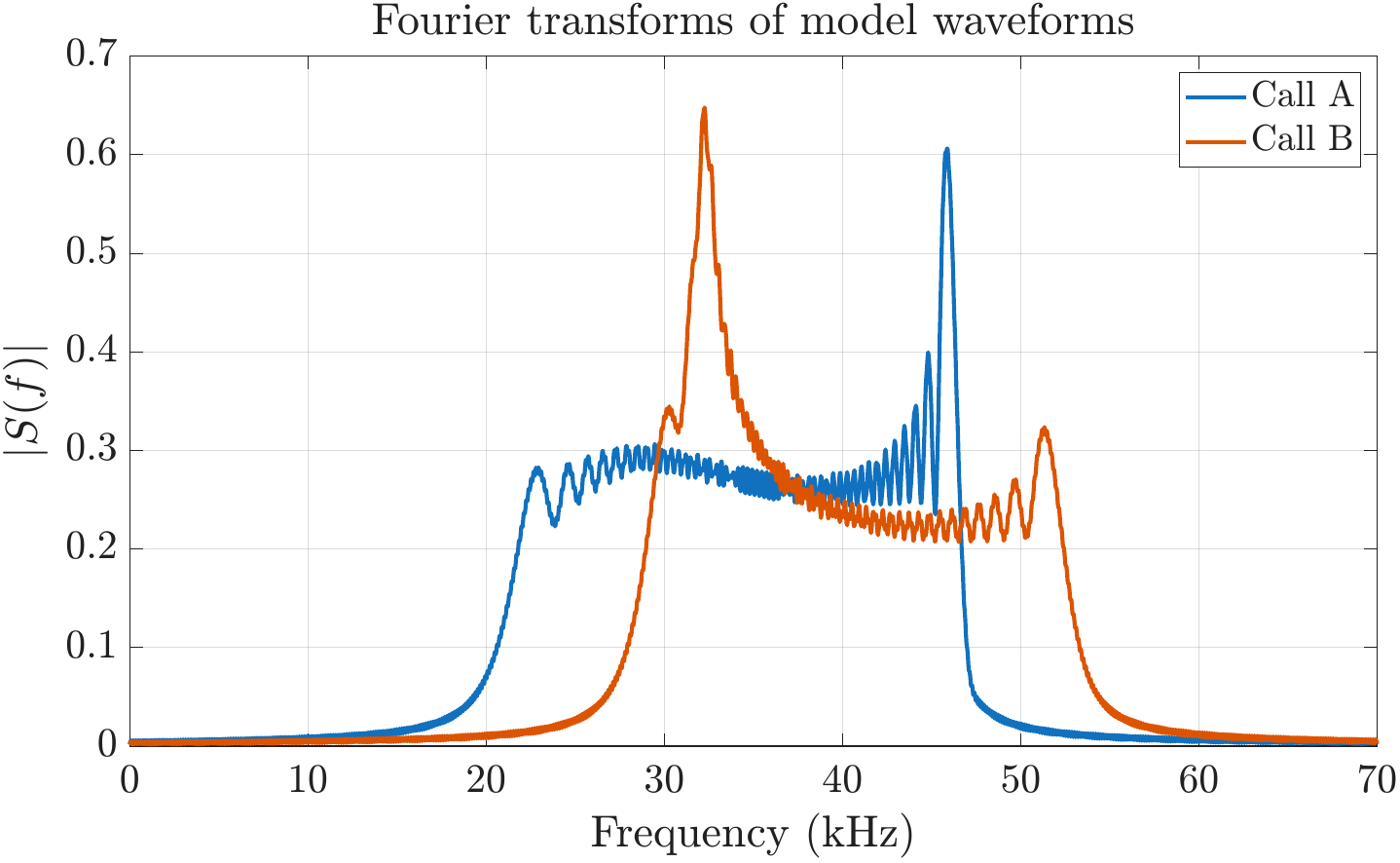} {8cm} {(b)}  }
\caption{(a) Fourth-order Chebyshev series fit for the instantaneous frequency for CPSFM models for two calls. Each call is about 7.5ms in duration and sweeps through roughly an octave of bandwidth.  Coefficients of the resulting Chebyshev series are $\{a_n\} = $\{34.9629  -13.1297  -1.0060  0.6347  -0.6599\} and $\{b_n\} = $\{38.8165  -11.9987  2.6602  -0.2573  -1.0253\}. (b) Spectra of the two model waveforms, computed via M-GBF expansions.}\label{CPSFM_Fig_4}
\end{figure*}

In sonar and radar applications, the autocorrelation function (ACF) indicates the ideal detection and range-finding performance of a system which transmits that waveform \cite{Ricker:2012}. Figure \ref{CPSFM_Fig_5} shows the ACF of each CPSFM waveform.   Also shown is the CCF of Call A with Call B.  This plot illustrates the degradation in detection and ranging performance that the sonar of individual A would encounter in the presence of the transmissions of individual B.  Call B is effectively an in-band ``sonar jammer'', but subtle differences in the shape of its FMF provide almost 12dB of jamming rejection for individual A.  Also in Fig.~\ref{CPSFM_Fig_5} is the CCF of Call A with pseudo-random noise with the same spectrum as Call A, as a reference.  This result suggests that subtle differences in spectro-temporal shape may facillitate jamming avoidance, allowing individual bats to correctly assign echoes even in the presence of other echolocating bats.
\begin{figure*}[h!]
\figline{ \fig{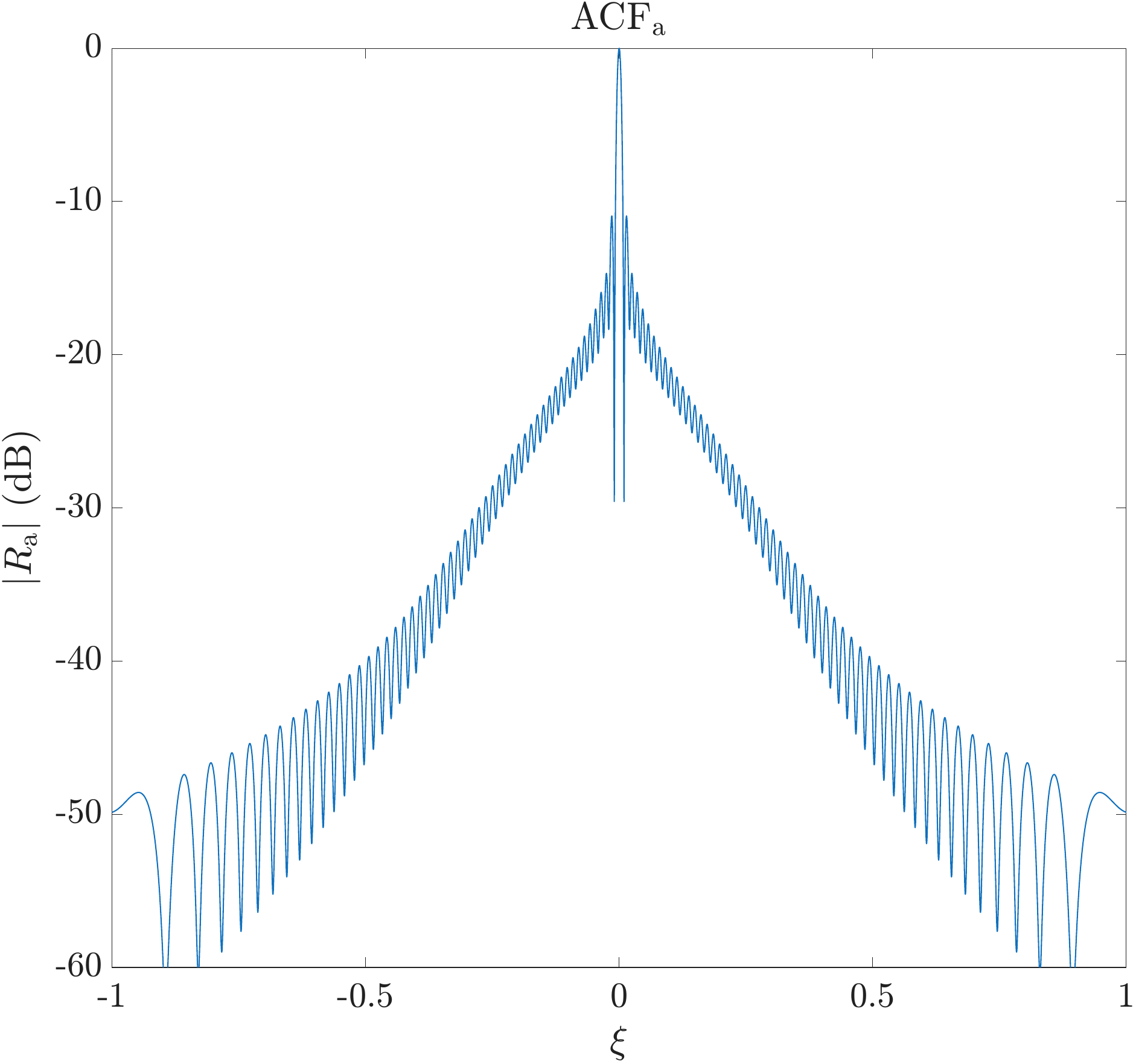} {8cm} {(a)}
	\fig{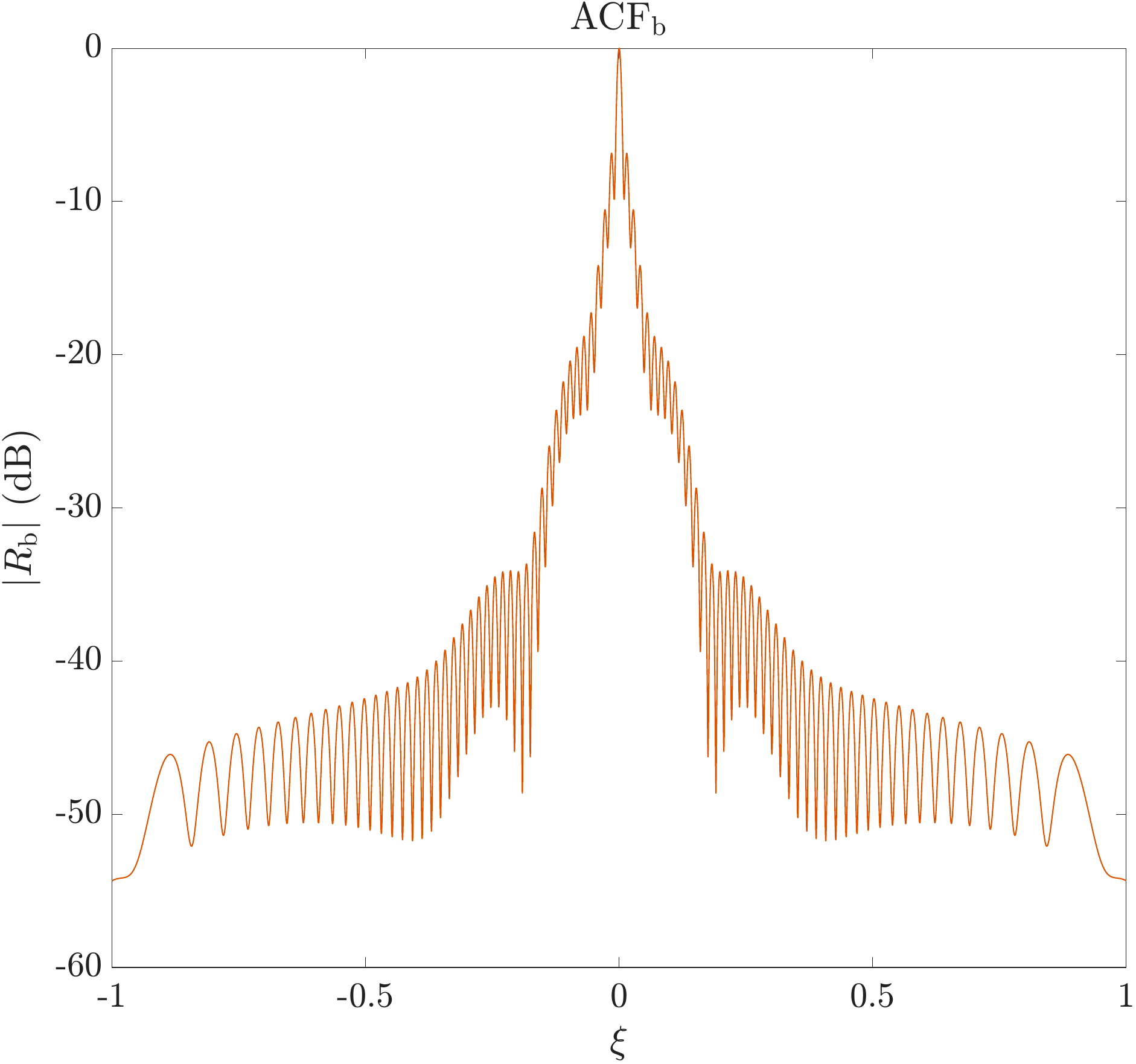} {8cm} {(b)}}
\figline{ \fig{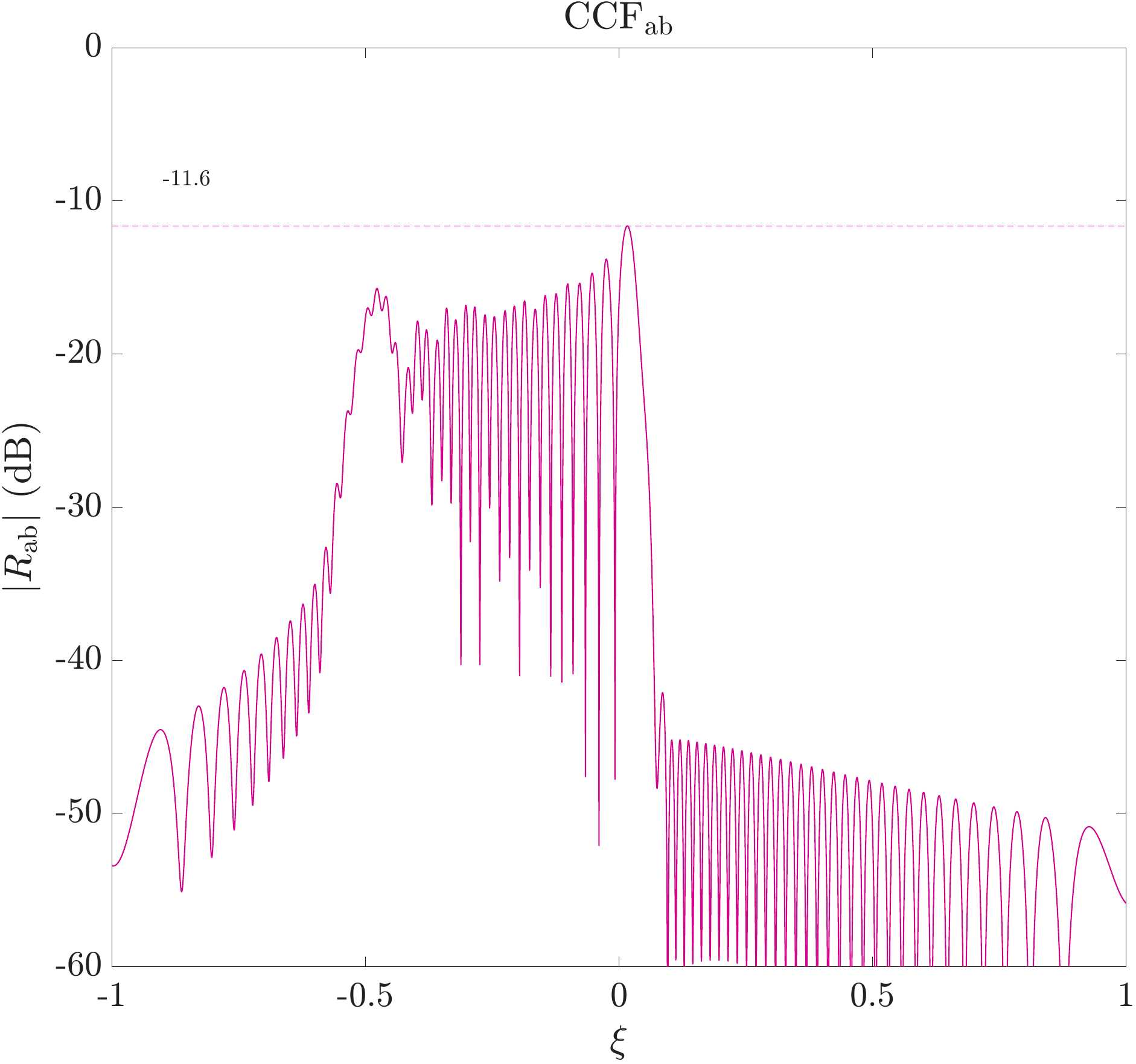} {8cm} {(c)}
	\fig{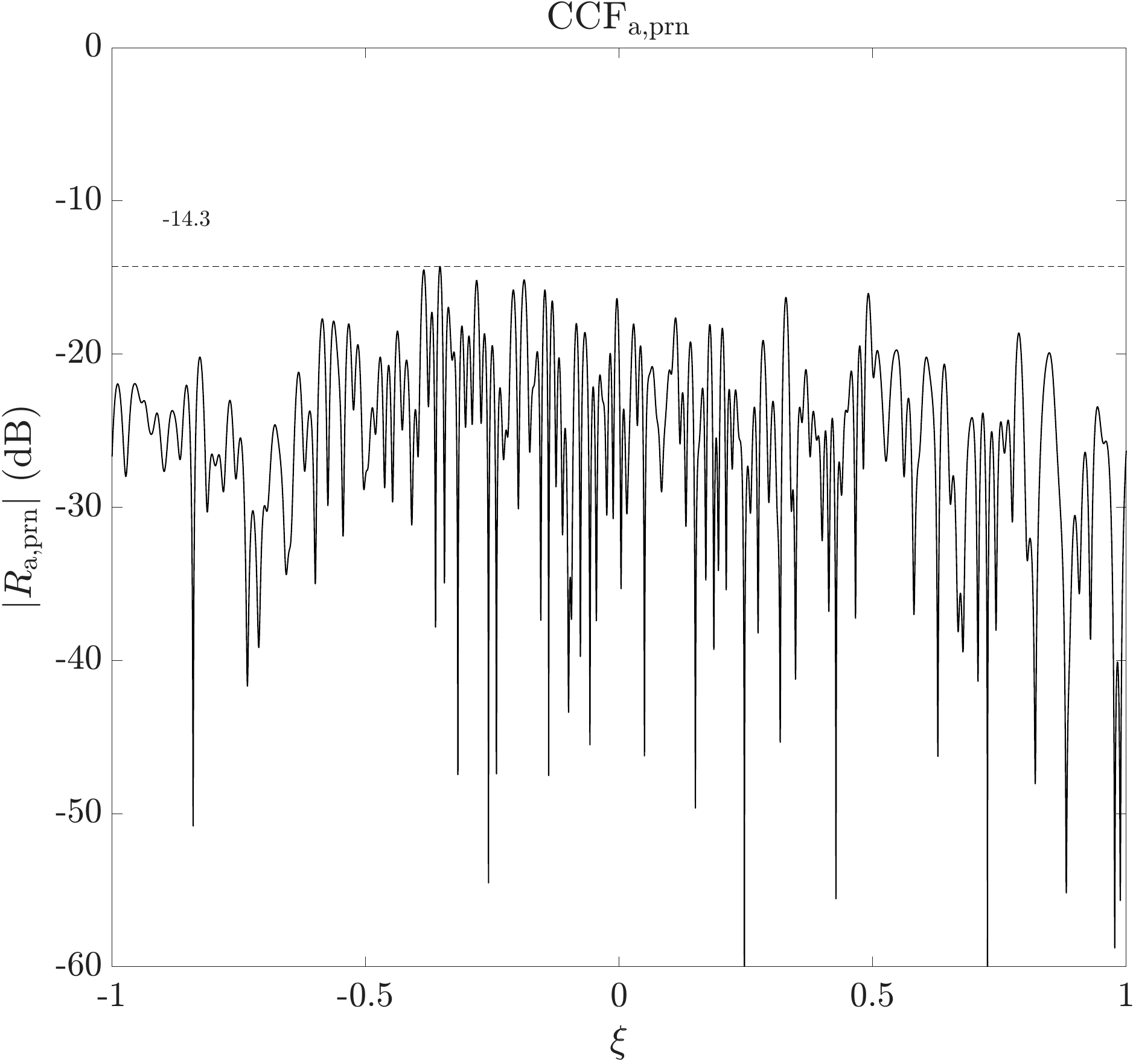} {8cm} {(d)}}
\caption{Auto- and cross-correlation functions of two model waveforms.  CCF peak is 11.6 dB attenuated relative to ACF, and is much less compact.  Lower right plot shows the CCF of waveform (a) versus pseudo-random noise whose spectrum magnitude is the same as that of waveform (a), which shows a 14.3 dB attenuation in peak response. } \label{CPSFM_Fig_5}
\end{figure*}
\begin{figure*}[h!]
\figline{ \fig{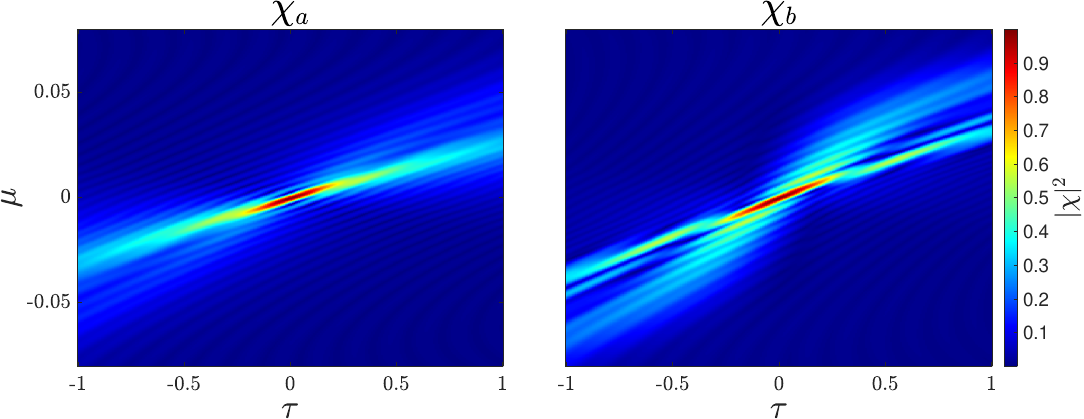} {16cm} {} \label{CPSFM_Fig_6}}
\caption{Ambiguity functions of two CPSFM model waveforms, caclulated according to Eq. \ref{AFboxed}. }
\end{figure*}
\\Figure \ref{CPSFM_Fig_6} shows the ambiguity functions of $s_a$ ans $s_b$, calculated according to Eq. \ref{spectrum1}. These two waveforms have significantly different Doppler tolerance, with $s_b$ exhibiting degraded range resolution and range ambiguity at even modest relative velocity. 
\fi
\section{Conclusion}
This paper presented a reformulation of the PPS model, introducing the CPSFM waveform model for finite-duration waveforms.  CPSFM waveforms are characterized by modulation functions which are Chebyshev series, and this format was shown as mathematically equivalent to an expansion in modified generalized Bessel functions.  This expansion enables the derivation of analytic solutions for Fourier transform, ambiguity function, and auto- and cross-correlation functions of an arbitrary-order CPSFM, and therefore an arbitrary-order PPS of finite duration.\\
Also presented was a waveform-design example showing a CPSFM approximation of an HFM waveform, including a quantitave exploration of the accuracy of this approximation relative to CPSFM order.  While application of the model to waveform design is beyond the scope of the present work, it seems likely that study and development of the model from an engineering perspective would be fruitful.  \\ 
A second example showed how the CPSFM model may be used in the analysis of echolocation signals of bats, facillitating the quantitative evaluation of active sonar performance, especially in the presence of jammers.  This application will be explored in more detail in future work.
%
\begin{acknowledgments}
	The authors gratefully acknowledge the contributions of David A. Hague, whose suggestion to investigate Chebyshev series was critically important, and whose expertise and guidance regarding generalized Bessel functions were invaluable.  The authors also thank Greg Allen and Terry Henderson for their generous efforts reviewing numerous drafts. L.K. acknowledges support from National Science Foundation Award \#2034885 and an Office of Naval Research Award N000141410588. A.T. was supported by National Science Foundation Postdoctoral Research Fellowship in Biology Award \#2410301.
\end{acknowledgments}
\section*{Author declarations}
\subsection*{Conflict of interest}
\noindent The authors have no conflicts to disclose.
 
\section*{Data availability}
\noindent The data that support the findings of this study are available from the corresponding author upon reasonable request.
\appendix
\section{M-GBF calculation} \label{appendix:M-GBF_calc}
Modified generalized Bessel functions play the role of coefficients in generalized Jacobi-Anger expansions:
\begin{equation} \label{jax}
\begin{split}
	\exp{\left( j \sum_{n=1}^N \alpha_n \cos{n\theta}\right)} &= \sum_{m=-\infty}^{\infty} \mathcal{I}_m(j \{\alpha_n\}) \, e^{jm\theta} ,
\end{split}
\end{equation}
where $\mathcal{I}_m$ is the modified generalized Bessel funtion of order $m$.\\
One approach to calculating these coefficients follows the integral formulation of the definition of the M-GBF:
\begin{equation} \label{MGBFInt}
	\mathcal{I}_m(j \{\alpha_n\}) = \frac{1}{\pi} \int_0^\pi \cos{m\theta} \, \exp{\left( j \sum_{n=1}^N  \alpha_n \cos{n\theta}\right)}\,d\theta .
\end{equation}
\\
Practical application of Eq.~\ref{MGBFInt} involves numerical integration in which the finite step size of $\theta$ must decrease with increasing $m$.\\
Another approach uses a set of recursive equations which relate M-GBFs to MBFs \cite{Dattoli:1992aa}:
\begin{equation} \label{MGBF_recursion}
\begin{split}
	\mathcal{I}_m(j\{\alpha_n\}) &\equiv \mathcal{I}_m(j\alpha_1, j\alpha_2,\ldots,j\alpha_N) \\
	&= \sum_{k=-\infty}^{\infty} \mathcal{I}_{m-Nk}(j\alpha_1, j\alpha_2,\ldots,j\alpha_{N-1}) \, I_{k}(j\alpha_N) ,
\end{split}
\end{equation}
where $I_k$ is the (single-variable) modified Bessel function of order $k$.\\
Since $\mathcal{I}_{\epsilon}$ is symmetric about zero with repect to integer order $\epsilon$, so that $\mathcal{I}_{-\epsilon}(j\{\alpha\}) = \mathcal{I}_{\epsilon}(j\{\alpha\})$, Eq.~\ref{MGBF_recursion} is, equivalently,
\begin{equation}
\begin{split}
	\mathcal{I}_m(j\{\alpha_n\}) = \sum_{k=-\infty}^{\infty} \mathcal{I}_{Nk-m}(j\alpha_1, j\alpha_2,\ldots,j\alpha_{N-1}) \, I_{k}(j\alpha_N) .
\end{split}
\end{equation}
For example, computation of the M-GBF coefficients for an order-4 CPSFM, whose phase modulation function is $\varphi(x) = \sum_{n=1}^5 \alpha_n T_n(x) = \sum_{n=1}^5 \alpha_n \cos{n \theta}$ follows the recursion:
\begin{equation*}
\begin{split}
	\mathcal{I}_m(\alpha_1) &= I_m(\alpha_1) \\
	\mathcal{I}_m(\alpha_1, \alpha_2) &= \sum_{k=-\infty}^{\infty} I_{2k-m}(\alpha_1) \, I_{k}(\alpha_2) \\
	\mathcal{I}_m(\alpha_1, \alpha_2, \alpha_3) &= \sum_{k=-\infty}^{\infty} \mathcal{I}_{3k-m}(\alpha_1, \alpha_2) \, I_{k}(\alpha_3) \\
	\mathcal{I}_m(\alpha_1, \alpha_2, \alpha_3, \alpha_4) &= \sum_{k=-\infty}^{\infty} \mathcal{I}_{4k-m}(\alpha_1, \alpha_2, \alpha_3) \, I_{k}(\alpha_4) \\
	\mathcal{I}_m(\alpha_1, \alpha_2, \alpha_3, \alpha_4, \alpha_5) &= \sum_{k=-\infty}^{\infty} \mathcal{I}_{5k-m}(\alpha_1, \alpha_2, \alpha_3, \alpha_4) \, \, I_{k}(\alpha_5) .\\
\end{split}
\end{equation*}
\section{Derivation of analytic solutions} \label{appendix:sp_derivations}
\subsection{Fourier transform}
The continuous Fourier transform of a signal $s(x)$ is
\begin{equation} \label{FTa0}
	S\big(f\big) = \mathcal{F}\{ s(t)\} = \infintegral s(t) \, e^{-j2 \pi f t} \, dt .
\end{equation}
For a finite-duration signal, such as $s(t) = \operatorname{rect}{(\frac{t}{T})}e^{\varphi(t)}$, the spectrum may be written
\begin{equation} \label{FTa1}
	S\big(f\big) = \int_{-T/2}^{T/2} s(t) \, e^{-j2 \pi f t} \, dt .
\end{equation}
If $s$ is a CPFSM waveform and we choose normalized time and frequency $x = \frac{2 t}{T}$ and $g = \frac{f}{2 T}$, so that $-1 \le x \le 1$, the resulting expression is
\begin{equation} \label{FTa2}
\begin{split}
	S\big(g; \{ \alpha_n\}\big) &= \symintegral{1} s(x) \, e^{-j2 \pi g x} \, dx\\
	&= \symintegral{1} \exp{\left(\summation{n}{1}{N} j \alpha_n T_n(x)\right)} \, e^{-j2 \pi g x} \, dx \\
	&= \symintegral{1} \exp{\left( -j 2 \pi g x + \summation{n}{1}{N} j \alpha_n T_n(x)\right)} \, dx .
\end{split}
\end{equation}
The kernel of the transform, $e^{j 2 \pi g x}$, may be absorbed into the summation, modifying the Chebyshev series slightly:
\begin{equation} \label{FTa2_1}
\begin{split}
	S\big(g; \{ \alpha_n\}\big) &=  \symintegral{1} \exp{\left( \summation{n}{1}{N} j \hat{\alpha}_n(g) T_n(x)\right)} \, dx ,
\end{split}
\end{equation}
where
\begin{align}
\begin{split} 
	\hat{\alpha}_n(g)
		&= \begin{cases}
			\alpha_n - j 2 \pi g, & n=1 \\
			\alpha_n, & n>1 .
		\end{cases} 
\end{split}
\end{align}
With a change of variables $x = \cos{\theta}$, we have:
\begin{equation} \label{FTa3}
\begin{split}
	S\big(g; \{ \alpha_n\}\big) &= \int_{\pi}^{0} {s(\cos{\theta}) \, e^{-j2\pi g \cos{\theta}} \, d(\cos{\theta}) } \\
	&= \int_{0}^{\pi} \exp{\left(\sum_{n=1}^{N} j \hat{\alpha}_n (g) \cos{n \theta}\right)} \, \sin{\theta} \, d\theta .
\end{split}
\end{equation}
The exponential term of the integrand may be re-written as an expansion in M-GBFs, most of which may be factored out of the integral:
\begin{equation} \label{FTa4}
\begin{split}
	S\big(g; \{ \alpha_n \}\big) &= \sum_{m=-\infty}^{\infty} \mathcal{I}_m(j \{\hat{\alpha}_n (g) \}) \int_0^{\pi} {  \, e^{jm\theta}  \, \sin{\theta} \, d\theta } .
\end{split}
\end{equation}
\newline
The integral in Eq.~\ref{FTa4} resolves to 
\begin{equation}  \label{gamma_integral}
\begin{split}
	\int_0^{\pi} {  \, e^{j m \theta} \, \sin{\theta} \, d\theta } &= \frac{e^{j m \theta} \left( \cos{\theta} - j m \sin{\theta} \right)}{m^2 - 1} \, \Bigg|_0^\pi\\
	&= \frac{e^{j m \pi} + 1}{m^2 - 1}.
\end{split}
\end{equation}
The cases $m = \pm 1$ are determined via L'Hopital's rule to be $\pm j \frac{\pi}{2}$. \\
To simplify the notation, we may define
\begin{equation*}
\begin{split}
	\gamma_{m} &\equiv \frac{(-1)^{ m } + 1}{1 -  m^2} \quad ,
\end{split}
\end{equation*}
where it is observed that
\begin{align}
\begin{split} 
	\gamma_{m}
		&= \begin{cases}
			\frac{2}{1 - m^2}, &  m  \operatorname{even} \\
			\pm j \frac{\pi}{2}, &   m  = \pm 1 \\
			0, &  m \text{ otherwise odd} .
		\end{cases} 
\end{split}
\end{align}
\\
Equation \ref{FTa2} is thus
\begin{equation} \label{spectrum1a}
	S\big(g; \{ \alpha_n\}\big) = \sum_{m=-\infty}^{\infty} \gamma_{m} \, \mathcal{I}_m(j \{\hat{\alpha}_n (g) \}) .
\end{equation}
\subsection{Cross-Ambiguity and Auto-Ambiguity functions}
The wideband cross-ambiguity function (CAF) of two unit-modulus signals $s_a(t)$ and $s_b(t)$ is
\begin{equation} \label{CAFa_definition}
\begin{split}
	\chi_{ab}(\tau, \nu) &= \frac{\sqrt{\nu}}{T} \int_{-\infty}^{\infty} s_a^*(t) \, s_b(\nu (t + \tau)) dt ,\\
\end{split}
\end{equation}
where $\nu = \frac{1+\mu}{1-\mu}$, $\mu = \frac{v}{c}$ (Mach index for sound), $c$ is propagation speed, and $v$ is relative velocity.  Normalizing by $\frac{\sqrt{\nu}}{T}$, where $T$ the waveform duration, enforces $|\chi(\tau, \nu)| \le 1$. \\
If $s_a$ and $s_b$ are CPSFM waveforms of identical duration, we use normalized time $x = \frac{2 t}{T}$ and $\xi = \frac{2 \tau}{T}$, so that
\begin{equation} \label{CAF_CPSFMa_0}
\begin{split}	
	\chi_{ab}(\xi, \nu) &= \frac{\sqrt{\nu}}{2} \int_{-\infty}^{\infty} s_a^*(x) \, s_b(\nu (x + \xi)) dx, 
\end{split}
\end{equation}
and the restrictions $-1 \le x \le 1$ and $-1 \le \nu(x + \xi) \le 1$ impose finite limits on the integral:
\begin{equation} \label{CAF_CPSFMa_1}
\begin{split}	
	\chi_{ab}(\xi, \nu) &= \frac{\sqrt{\nu}}{2} \int_{x_1}^{x_2} s_a^*(x) \, s_b(\nu (x + \xi)) dx ,
\end{split}
\end{equation}
where
\begin{equation} \label{xi_limits_AFa}
	\begin{split}
		x_1 &= \max{(-1, -\frac{1}{\nu} -\xi)} \\
		x_2 &= \min{(1, \frac{1}{\nu} -\xi)}.
	\end{split}
\end{equation}
\\
From the CPSFM definition Eq.~\ref{tdw}
\begin{equation} \label{CAFa_s0}
\begin{split}	
	s_a^* \big( x \big) &= \exp{ \left( -j \, \varphi_a(x) \right)} \\
	s_b \big( \nu [x + \xi] \big) &= \exp{ \left( j \, \varphi_b( \nu [x + \xi]) \right)} ,
\end{split}
\end{equation}
where $\varphi_a$ and $\varphi_b$ are such that
\begin{equation} \label{CAFa_s}
\begin{split}	
	s_a^* \big( x \big) &= \exp{ \left( -j \, \sum_{n=1}^N \alpha_n T_n \big( x \big) \right)} \\
	s_b\big( \nu [x + \xi]x \big) &= \exp{\left( j \, \sum_{n=1}^N \beta_n T_n \big( \nu [x + \xi] \big) \right)}.
\end{split}
\end{equation}
According to Eq.~\ref{cps_scaleshift1}, the Chebyshev series in the second equation in Eq.~\ref{CAFa_s} may be written
\begin{equation} \label{phi_hat_a}
	 \tilde{\beta}_0 + \sum_{n=1}^N \tilde{\beta}_n T_n (x) = \sum_{n=1}^N \beta_n T_n \big( \nu [x + \xi] \big) ,
\end{equation}
where $\tilde{\beta}_0$ represents a phase shift induced by $\xi$-shifting and $\nu$-scaling $s_b$, which is generally nonzero.
Now Eq.~\ref{CAF_CPSFMa_1} becomes
\begin{equation}
\begin{split}
	\chi&(\xi, \nu ; \{ \alpha_n, \beta_n\}) = \rho_{\xi, \nu} \, \frac{\sqrt{\nu}}{2} \\ 
	\times &\int_{x_1}^{x_2}  \, \exp{ \left( j \sum_{n=1}^{N} \big( \tilde{\beta}_n - \alpha_n \big) \, T_n(x)\right)} dt ,
\end{split}
\end{equation}
where the phase offset $\rho_{\xi, \nu} = \exp{ \left( j \tilde{\beta}_{0} \right)}$.
Limits $x_1, x_2$ are defined in Eq.~\ref{xi_limits_AFa}, and the parameterization indicates the dependence on $\{ \alpha_n \}$ and $\{ \beta_n \}$.\\
A change of variables: $x = \cos{\theta}, \quad T_n(\cos{\theta}) = \cos{n\theta}$ produces
\begin{equation}
\begin{split}
	\chi_{ab} & (\xi, \nu ; \{ \alpha_n, \beta_n\}) = \rho_{\xi, \nu} \, \frac{\sqrt{\nu}}{2} \\
		& \times \int_{\theta_2} ^{\theta_1} \exp{ \left( j \sum_{n=1}^{N} \big( \tilde{\beta}_n - \alpha_n \big) \, \cos{n\theta}\right)} \, \sin{\theta} \, d\theta ,
\end{split}
\end{equation}
where $\theta_{i} = \operatorname{acos} x_{i}$, $i \in \{1,2\}$.   (Note $0 \le \theta_2 < \theta_1 \le \pi$.)\\
Most of the integrand is suitable for Jacobi-Anger expansion in modified generalized Bessel functions:
\begin{equation}
\begin{split}
	\chi_{ab}&(\xi, \nu ; \{ \alpha_n, \beta_n\}) = \rho_{\xi, \nu} \, \frac{\sqrt{\nu}}{2} 
	\sum_{m=-\infty}^{\infty}\mathcal{I}_m \Big(j  \big\{ \tilde{\beta}_n - \alpha_n \big\} \Big)
	 \int_{\theta_2}^{\theta_1} e^{j m \theta} \sin{\theta} \, d\theta .
\end{split}
\end{equation}
We define $\gamma_m(\theta_1, \theta_2)$ to be the integral multiplicand, which resolves to
\begin{equation}
\begin{split}
	\gamma_m&(\theta_1, \theta_2) = \gamma_m\left(\acos{x_1}, \acos{x_2}\right) \equiv \int_{\theta_2} ^{\theta_1} e^{j m \theta} \sin{\theta} \, d\theta \\
	&= 	\frac{1}{m^2-1} \Big[ e^{j m \theta_1} ( \cos{\theta_1} - j m \sin{\theta_1}) \\
	  & \quad - e^{j m \theta_2}( \cos{\theta_2} - j m \sin{\theta_2})\Big] \quad , m \ne \pm 1 .
\end{split}
\end{equation}
The cases $m = \pm 1$ are determined via L'Hopital's rule to be \\
\begin{equation}
\begin{split}
	\gamma_{\pm 1}(\theta_1, \theta_2) &= \frac{1}{4} ( 1 - e^{j 2 m \theta} + j 2 m \theta) \Bigg|_{\theta_2} ^{\theta_1}\\
	&= \frac{1}{4} \Big[ e^{j 2 m \theta_2} - e^{j 2 m \theta_1} + j 2 m (\theta_1 - \theta_2) \Big] .
 \end{split}
\end{equation}
Finally, 
\begin{equation} \label{CAFa_boxed}
	\begin{split}
	\chi_{ab}&(\xi, \nu ; \{ \alpha_n, \beta_n\}) = \rho_{\xi, \nu} \, \frac{\sqrt{\nu}}{2} 
	 \sum_{m=-\infty}^{\infty} \gamma_m(\theta_1, \theta_2) \, \mathcal{I}_m \Big(j \{ \tilde{\beta}_n - \alpha_n\} \Big)
	\end{split}
\end{equation}
The auto-ambiguity function, which we abbreviate simply AF, is the cross-ambiguity function when $s_a(x) = s_b(x)$:
\begin{equation} \label{AFa_boxed}
	\begin{split}
	\chi&(\xi, \nu ; \{ \alpha_n\}) = \rho_{\xi, \nu} \, \frac{\sqrt{\nu}}{2} 
	 \sum_{m=-\infty}^{\infty} \gamma_m(\theta_1, \theta_2) \, \mathcal{I}_m \Big(j \{ \tilde{\alpha}_n - \alpha_n\} \Big) .
	\end{split}
\end{equation}
%



\bibliography{CPSFM_1c.bib}+

@article{kuklinski2021identities,
	archiveprefix = {arXiv},
	author = {Parker Kuklinski and David A. Hague},
	eprint = {1908.11683},
	journal = {arXiv preprint arXiv:1908.11683},
	primaryclass = {id='math.GM' full_name='General Mathematics' is_active=True alt_name=None in_archive='math' is_general=True description='Mathematical material of general interest, topics not covered elsewhere'},
	title = {Identities and Properties of Multi-Dimensional Generalized {B}essel Functions},
	year = {2021}}

@inproceedings{davies2024,
	author = {Davies, Jonathan and Dinale, Justin and Mitchell, Paul and Dol, Henry and Van Walree, Paul and Nissen, Ivor},
	bdsk-color = {3},
	booktitle = {2024 Seventh Underwater Communications and Networking Conference (UComms)},
	organization = {IEEE},
	pages = {1--4},
	title = {Towards Open Underwater Acoustic Channel Models \& Tools for Acoustic Communication ({ACOMMS}) Waveform-Receiver Performance Assessment},
	year = {2024}}

@article{kloepper2024hawkear,
	author = {Kloepper, L and Taylor, G and Domski, P and Vanderelst, D and Eveland, K and Stevenson, R},
	bdsk-color = {3},
	journal = {Methods in Ecology and Evolution},
	number = {7},
	title = {Hawk{E}ar: a bird-borne visual and acoustic platform for eavesdropping the behavior of mobile animals},
	volume = {15},
	year = {2024}}

@book{Ricker:2012,
	author = {Ricker, Dennis W},
	bdsk-color = {3},
	publisher = {Springer Science \& Business Media},
	title = {Echo signal processing},
	volume = {725},
	year = {2012}}

@article{Scaglione1998,
	author = {Scaglione, A. and Barbarossa, S.},
	bdsk-color = {3},
	journal = {IEEE Signal Processing Letters},
	number = {9},
	pages = {237-240},
	title = {On the spectral properties of polynomial-phase signals},
	volume = {5},
	year = {1998}}

@article{Canny1986,
	author = {Canny, John},
	bdsk-color = {3},
	journal = {Pattern Analysis and Machine Intelligence, IEEE Transactions on},
	month = {12},
	pages = {679 - 698},
	title = {A Computational Approach To Edge Detection},
	volume = {PAMI-8},
	year = {1986}}

@book{boyd2001chebyshev,
	author = {Boyd, John P},
	bdsk-color = {3},
	publisher = {Courier Corporation},
	title = {Chebyshev and Fourier Spectral Methods},
	year = {2001}}

@book{Bradbury:2011aa,
	address = {Sunderland, Mass},
	author = {Bradbury, J. W. and Vehrencamp, Sandra Lee},
	bdsk-color = {3},
	edition = {2nd},
	publisher = {Sinauer Associates},
	series = {Animal communication},
	title = {Principles of animal communication},
	year = {2011}}

@article{simmons1980acoustic,
	author = {Simmons, James A and Stein, Roger A},
	bdsk-color = {3},
	journal = {Journal of Comparative Physiology},
	number = {1},
	pages = {61--84},
	title = {Acoustic imaging in bat sonar: echolocation signals and the evolution of echolocation},
	volume = {135},
	year = {1980}}

@article{Vespe:2009aa,
	author = {Vespe, Michele and Jones, Gareth and Baker, Chris J.},
	bdsk-color = {3},
	journal = {IEEE Signal Processing Magazine},
	number = {1},
	pages = {65--75},
	title = {Lessons for Radar: {W}aveform diversity in echolocating mammals},
	volume = {26},
	year = {2009}}

@article{Djurovic:2017aa,
	author = {Djurovi{\'c}, Igor and Simeunovi{\'c}, Marko and Wang, Pu},
	bdsk-color = {3},
	journal = {Signal Processing},
	pages = {48--66},
	title = {Cubic phase function: A simple solution to polynomial phase signal analysis},
	volume = {135},
	year = {2017}}

@book{Cook_Bernfeld_1967,
	address = {New York},
	author = {Cook, Charles E. and Bernfeld, Marvin.},
	bdsk-color = {3},
	publisher = {Academic Press},
	series = {Electrical science series},
	title = {Radar signals: an introduction to theory and application},
	year = {1967}}

@article{Dattoli:1992aa,
	author = {Dattoli, G. and Mari, C. and Torre, A. and Chiccoli, C. and Lorenzutta, S. and Maino, G.},
	bdsk-color = {3},
	journal = {Journal of Scientific Computing},
	number = {2},
	pages = {175--196},
	title = {Analytical and numerical results on {M}-variable generalized {B}essel functions},
	volume = {7},
	year = {1992}}

@article{YangJ.2006Dpoh,
	author = {Yang, J. and Sarkar, T. K.},
	bdsk-color = {3},
	journal = {Microwave and Optical Technology Letters},
	number = {6},
	pages = {1174-1179},
	title = {Doppler-invariant property of hyperbolic frequency modulated waveforms},
	volume = {48},
	year = {2006}}

@article{PelegS.1991Eaco,
	author = {Peleg, S. and Porat, B.},
	bdsk-color = {3},
	journal = {IEEE Transactions on Information Theory},
	number = {2},
	pages = {422-430},
	title = {Estimation and classification of polynomial-phase signals},
	volume = {37},
	year = {1991}}

@article{GershmanA.B.2001Epom,
	author = {Gershman, A.B. and Pesavento, M. and Amin, M.G.},
	bdsk-color = {3},
	journal = {IEEE Transactions on Signal Processing},
	number = {12},
	pages = {2924-2934},
	title = {Estimating parameters of multiple wideband polynomial-phase sources in sensor arrays},
	volume = {49},
	year = {2001}}

@book{Trefethen_2020,
	address = {Philadelphia},
	author = {Trefethen, Lloyd N.},
	bdsk-color = {3},
	edition = {Extended},
	publisher = {Society for Industrial and Applied Mathematics},
	series = {Other titles in applied mathematics ; 164},
	title = {Approximation theory and approximation practice},
	year = {2020}}

@article{Lorenzutta1995,
	author = {S. Lorenzutta and G. Maino and G. Dattoli and A. Torre and C. Chiccoli},
	bdsk-color = {3},
	journal = {Rendiconti di Matematica e delle sue Applicazioni (1981)},
	number = {3},
	pages = {405-420},
	title = {On infinite-variable {B}essel functions},
	volume = {15},
	year = {1995}}

@article{HagueDavidA.2017TGSF,
	author = {Hague, David A. and Buck, John R.},
	bdsk-color = {3},
	journal = {IEEE Journal of Oceanic Engineering},
	number = {1},
	pages = {109-123},
	title = {The Generalized Sinusoidal Frequency-Modulated Waveform for Active Sonar},
	volume = {42},
	year = {2017}}

@article{HagueDavidA.2021ATWD,
	author = {Hague, David A.},
	bdsk-color = {3},
	journal = {IEEE Transactions on Aerospace and Electronic Systems},
	number = {2},
	pages = {1274-1287},
	title = {Adaptive Transmit Waveform Design Using Multitone Sinusoidal Frequency Modulation},
	volume = {57},
	year = {2021}}

@article{Blunt_Mokole_2016,
	author = {Blunt, Shannon D. and Mokole, Eric L.},
	bdsk-color = {3},
	journal = {IEEE Aerospace and Electronic Systems Magazine},
	number = {11},
	pages = {2-42},
	title = {Overview of radar waveform diversity},
	volume = {31},
	year = {2016}}





\end{document}